\documentclass[secnumarabic,amssymb, nobibnotes, preprint]{revtex4-2}
\usepackage{graphicx}
\usepackage{caption}
\usepackage{subcaption}
\usepackage{array}

\begin{document}

\title{Power-Efficient Silicon Nitride Thermo-Optic Phase Shifters for Visible Light}

\author{Zheng Yong,$^{1,2,4,*}$ Hong Chen,$^{1,4}$ Xianshu Luo,$^{3}$ Alperen Govdeli,$^{1,2}$ Hongyao Chua,$^{3}$ Saeed S. Azadeh,$^{1}$ Andrei Stalmashonak,$^{1}$ Guo-Qiang Lo,$^{3}$ Joyce K.S. Poon,$^{1,2}$and Wesley D. Sacher$^{1,**}$}

\affiliation{$^{1}$Max Planck Institute of Microstructure Physics, Weinberg 2, 06120 Halle, Germany}
\affiliation{$^{2}$Department of Electrical and Computer Engineering, University of Toronto, 10 King’s College Road, Toronto, Ontario M5S 3G4, Canada}
\affiliation{$^{3}$Advanced Micro Foundry Pte. Ltd., 11 Science Park Road, Singapore Science Park II, 117685, Singapore} 
\affiliation{$^{4}$These authors contributed equally to the work}

\affiliation{$^{*}$zheng.yong@mail.utoronto.ca\\
$^{**}$wesley.sacher@mpi-halle.mpg.de}

\begin{abstract}
We demonstrate power-efficient, thermo-optic, silicon nitride waveguide phase shifters for blue, green, and yellow wavelengths. The phase shifters operated with low power consumption due to a suspended  structure and multi-pass waveguide design. The devices were fabricated on 200-mm silicon wafers using deep ultraviolet lithography as part of an active visible-light integrated photonics platform. The measured power consumption to achieve a $\pi$ phase shift (averaged over multiple devices) was 0.78, 0.93, 1.09, and 1.20 mW at wavelengths of 445, 488, 532, and 561 nm, respectively. The phase shifters were integrated into Mach-Zehnder interferometer switches, and $10- 90$\% rise(fall) times of about 570(590) $\mu$s were measured.
\end{abstract}

\maketitle

\section{Introduction}

Integrated photonics for visible wavelengths is becoming the focus of a growing number of research efforts in recent years \cite{Subramanian_2013,Romero_Garcia_OE_2013,Domenech_2018,sacher2019visible,Al2O3_JSTQE_2019,Mohanty_2020,laser_engine_OE_2021}. By adapting the tools and fabrication infrastructure developed for silicon (Si) photonics at telecommunication wavelengths, visible-light photonic circuits with $\sim$ 10 - 100's of components have been demonstrated for neurophotonics \cite{Mohanty_2020,Sacher_Neurophotonics_2021}, beam scanners \cite{poulton2017}, and quantum information \cite{Niffenegger_Nature_2020,Mehta_Nature_2020} applications. Fabrication on 200- and 300-mm diameter Si wafers has been enabled by the development of low-loss silicon nitride (SiN) \cite{Subramanian_2013,sacher2019visible} and aluminum oxide (Al$_2$O$_3$) \cite{west2019low,Al2O3_JSTQE_2019} waveguides. Despite the visible-light transparency of these materials and their compatibility with Si photonics fabrication \cite{timurdogan2019apsuny,wilmart2019versatile,sacher2018monolithically,absil2017reliable,towerjazz}, the development of efficient phase shifters based on these waveguide materials has remained challenging. Conventional SiN thermo-optic phase shifters have a power consumption to achieve a $\pi$ phase shift, $P_{\pi}$, of about 20-30 mW, owing to the relatively low thermo-optic coefficient of SiN of $2.45 \times 10^{-5}$ K$^{-1}$ \cite{Mohanty_2020,arbabi2013measurements}. The thermo-optic coefficient of Al$_2$O$_3$ ($2.75 \times 10^{-5}$ K$^{-1}$ in \cite{west2019low}) is similar to that of SiN. Phase shifters based on thermally-tuned SiN microdisks and microrings have $P_{\pi}$ of $0.68 - 2.1$ mW at wavelengths of 488 and 530 nm \cite{Liang_CLEO_2019,Huang_CLEO_2020,yu2021micron}; however, the low power consumption is attained at the expense of increased wavelength and fabrication error sensitivity compared to non-resonant structures. SiN phase shifters with liquid crystal cladding are expected to have significantly lower power consumption and such devices have been demonstrated at a wavelength of $\lambda = 630$ nm \cite{notaros2018integrated,notaros2019liquid}, but post-processing steps are required to apply and encapsulate the liquid crystal. Low-loss lithium niobate nanophotonic waveguides have been demonstrated at red and near-infrared wavelengths ($634 - 638$ nm, $720 - 850$ nm), and ultra-low power phase shifters have been demonstrated at a wavelength of 850 nm \cite{desiatov2019ultra}; however, the incompatibility of lithium niobate processing with standard silicon photonics fabrication limits the levels of integration that may be achieved.

Here, we present visible-light, power-efficient, SiN, thermo-optic phase shifters based on a suspended heater structure with folded waveguides. Suspended heaters enable the power-efficient generation of isolated hot spots on-chip \cite{Kasahara_ECOC_2008}, and the thermo-optic phase shift increases with the number of times a waveguide is passed through the suspended volume \cite{Densmore_OE_2009}.  As the structure is non-resonant, operation over a wide wavelength range is possible, and here, we demonstrate $P_{\pi} = $ 0.78, 0.93, 1.09, 1.20 mW at wavelengths of 445, 488, 532, and 561 nm. Efficient suspended and folded-waveguide thermo-optic phase shifters have been demonstrated with Si waveguides at telecommunication wavelengths in \cite{Kasahara_ECOC_2008,Densmore_OE_2009,fang2011ultralow,lu2015michelson,MurrayOE2015,celo2016thermo}; however, to our knowledge, this is the first demonstration of such a phase shifter in the visible spectrum. The power efficiency is comparable to microring/microdisk (resonant) thermo-optic phase shifters \cite{Liang_CLEO_2019,Huang_CLEO_2020} and a record among non-resonant visible-light SiN thermo-optic phase shifters. The phase shifters were fabricated on 200-mm diameter Si wafers as part of our visible-light integrated photonics platform, which also includes passive SiN waveguide devices \cite{sacher2019visible}, high-efficiency bi-layer edge-couplers \cite{yiding2021ec}, and SiN-on-Si photodetectors \cite{yiding2021pd}.

\section{Thermo-optic phase shifter design}
\label{sec:design}

\begin{figure}
\centering
    \subfloat[\centering ]{{\includegraphics[width=0.7\textwidth]{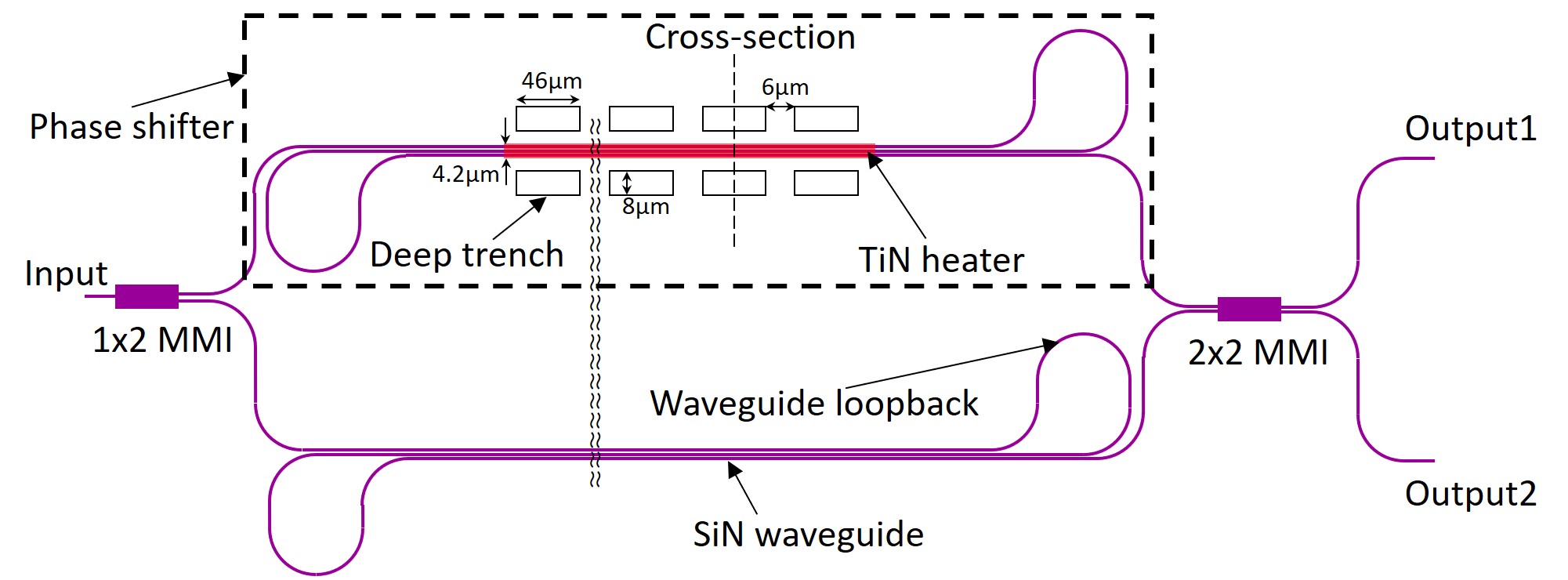} }}%
    \qquad
    \subfloat[\centering ]{{\includegraphics[width=0.4\textwidth]{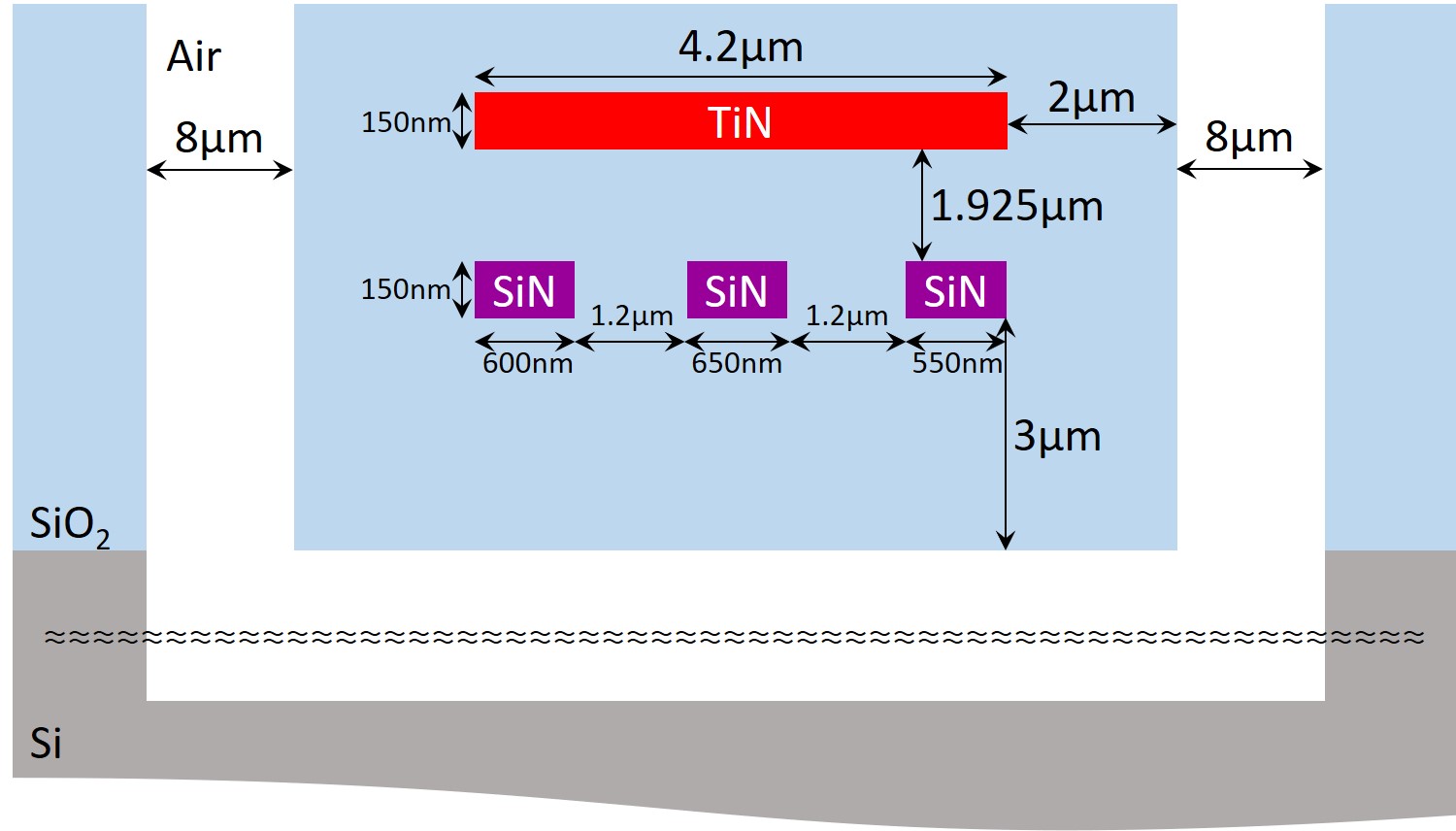} }}%
    \subfloat[\centering ]{{\includegraphics[width=0.3\textwidth]{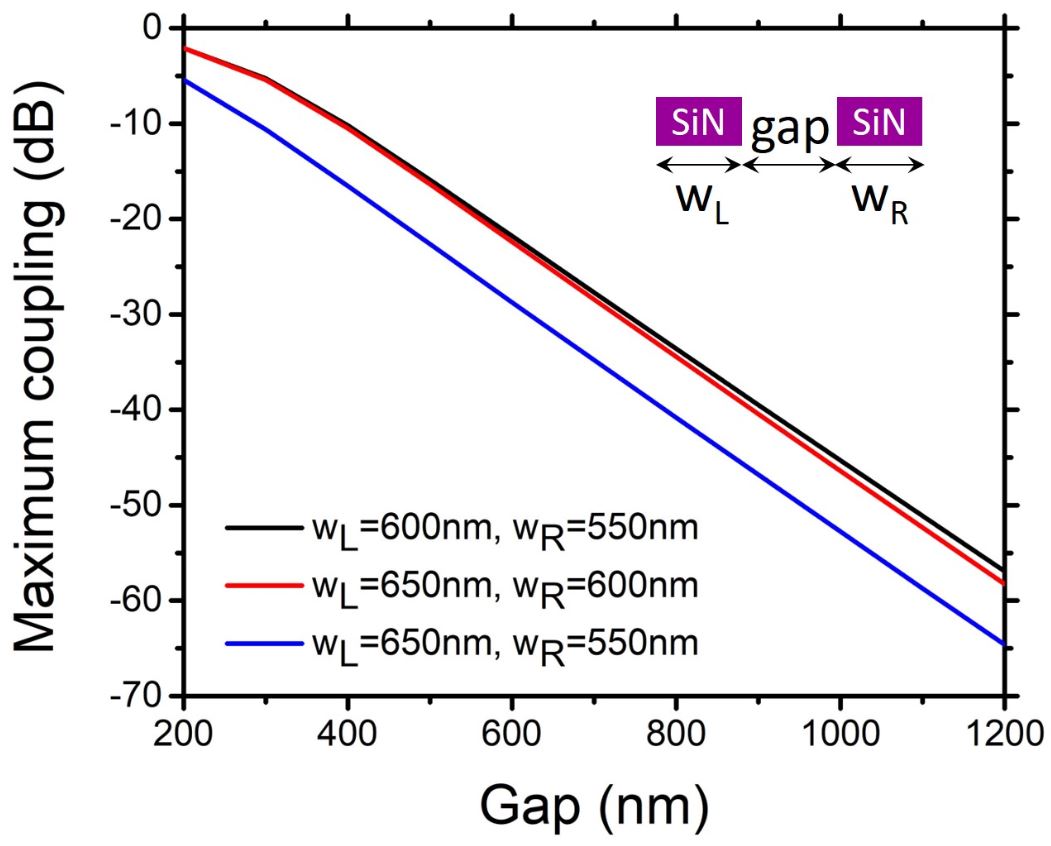} }}%
    \qquad
    \subfloat[\centering ]{{\includegraphics[width=0.7\textwidth]{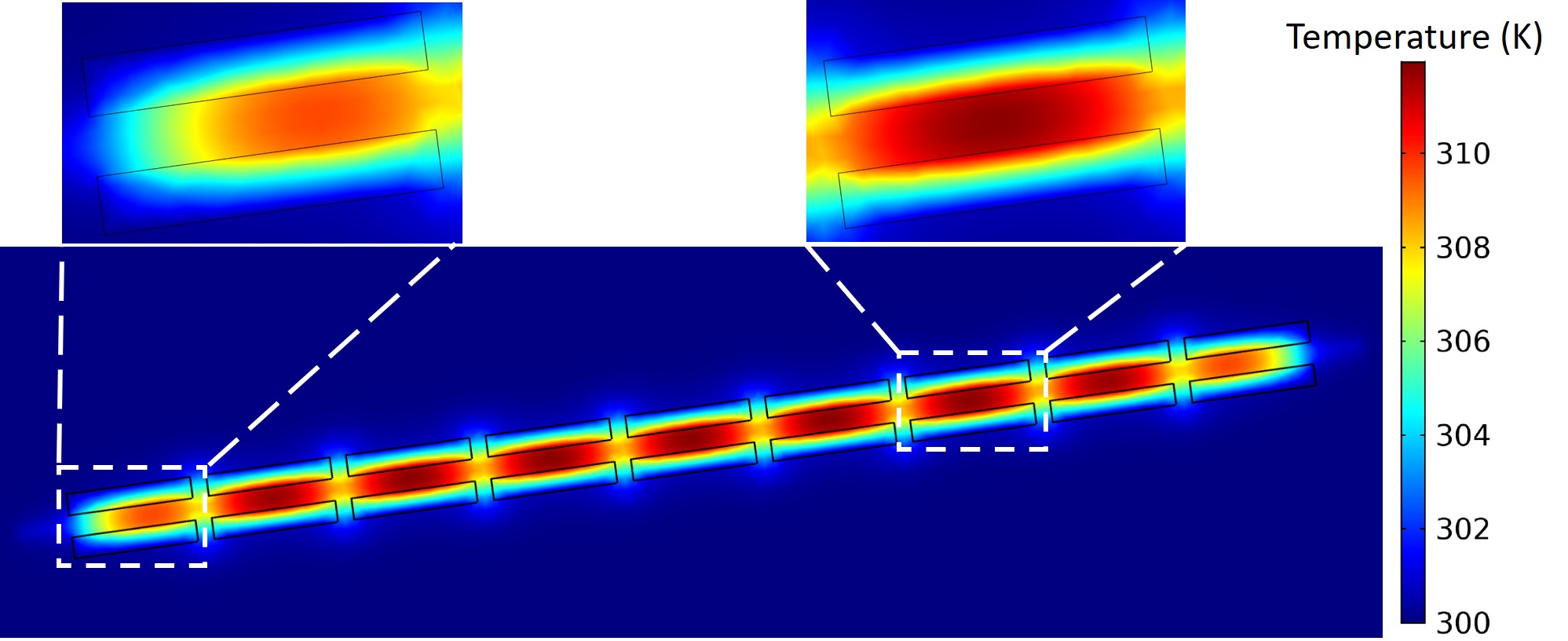} }}%
\caption{SiN thermo-optic phase shifter design. (a) Top-down schematic of the phase shifter integrated into a MZI test structure (not to scale). The input and output edge couplers are not included in the schematic. (b) Cross-section schematic of the suspended region of the phase shifter corresponding to the dashed line in (a) (not to scale); metal wiring and via layers not shown. (c) The simulated maximum coupling between adjacent waveguides in the phase shifter at a wavelength of 561 nm. (d) Three-dimensional thermal simulation of the temperature rise of the phase shifter suspended region when 1.19 mW power is dissipated by the TiN heater.}\label{fig:fig1}
\end{figure}

A schematic of the thermo-optic phase shifter is shown in Fig. \ref{fig:fig1}(a). To facilitate characterization, the phase shifter was integrated into a Mach-Zehnder interferometer (MZI) with $1\times2$ and $2\times2$ multimode interference (MMI) couplers and nominally identical waveguide paths in each arm. The devices were designed in our visible-light integrated photonics platform using the 150 nm thick SiN waveguide layer deposited by plasma enhanced chemical vapor deposition (PECVD); the refractive index of the SiN was reported in \cite{sacher2019visible}. 

The phase shifter consisted of a suspended SiO$_2$ bridge with a 500 $\mu$m long titanium nitride (TiN) heater and folded SiN waveguides. The suspended region was formed by undercut etching of the Si substrate via rows of deep trenches (9 trenches per row); the gaps between the trenches formed SiO$_2$ anchors for device robustness \cite{lu2015michelson,MurrayOE2015}. The SiN waveguides looped through the suspended region 3 times using a folded structure with 60 $\mu$m radius bends, and the total SiN waveguide length under the TiN heater was 1.5 mm.

The waveguide dimensions were designed for the transverse-electric (TE) polarization. The designed cladding thicknesses, Fig. \ref{fig:fig1}(b), ensured negligible TiN and Si substrate absorption; the computed TiN absorption loss at $\lambda = 561$ nm for a 550 nm wide waveguide was $< 10^{-6}$ dB/cm. Since $P_{pi}$ decreases with a narrower suspended region width, we applied the design strategy in \cite{MurrayOE2015} and selected dissimilar widths of 600, 650, and 550 nm for the 3 folded waveguides, such that the waveguide gap can remain small at 1.2 $\mu$m without significant inter-waveguide coupling. The design was informed by the simulations in Fig. \ref{fig:fig1}(c), which shows the simulated maximum coupling ratio between two waveguides at $\lambda = 561$ nm; the worst cross-coupling between the adjacent SiN waveguides is $<-55$ dB when the gap is 1.2 $\mu$m.

The thermo-optic phase shifter performance was modeled using COMSOL Multiphysics (for thermal simulations) and Lumerical MODE Solutions (for optical mode simulations). The simulations assumed a TiN resistance of 10 $\Omega$/sq, and thus, a heater resistance of 1190 $\Omega$. The undercut etch was modeled as elliptical channels in the Si substrate centered on the rows of deep trenches; the channels extended 8 $\mu$m laterally and 18 $\mu$m vertically from the edges of the deep trenches. The undercut etch was estimated from cross-section images of fabricated samples. The thermal simulations included a 2 mm thick air region above the device, and the bottom surface of the Si substrate was to set to a temperature of 300 K. The other boundaries of the simulated volume were set to be thermally insulating. The simulated temperature increase was highest in the suspended region, as shown in Fig. \ref{fig:fig1}(d). For an applied current of 1 mA, equivalent to an electrical power dissipation of 1.19 mW, the calculated average temperature increase of the SiN waveguides under the TiN heater was 8.9 K. Next, the phase shift of the SiN waveguide was calculated using
\begin{equation}
\Delta\phi=\frac{d n_{eff}}{dT}\Delta T \frac{2\pi}{\lambda}  L_{wg},
\end{equation}\label{eq:phaseshift}
where $\frac{dn_{eff}}{dT}$ is the change of the SiN waveguide effective index with temperature, $\Delta T$ is the temperature change, and $L_{wg} = 1.5$ mm is the length of SiN waveguide under the TiN heater. To relate $\Delta n_{eff}$ to the temperature change, we used the thermo-optic coefficients reported in \cite{arbabi2013measurements}; $2.45 \times 10^{-5}$ K$^{-1}$ for SiN and $0.95 \times 10^{-5}$ K$^{-1}$ for SiO$_2$, measured in the C band. The thermo-optic coefficients measured in \cite{Elshaari_IEEE_Photonics_Journal_2016} at wavelengths around 880 nm closely agree with these results. Via optical mode simulations, the calculated $\Delta n_{eff} = 1.94 \times 10^{-4}$ for $\Delta T = 8.9$ K at $\lambda = 445$ nm, equivalent to a phase shift of 1.31$\pi$. Therefore, the simulated $P_\pi$ of the phase shifter is 0.91 mW at 445 nm.

\begin{table}[ht!]
\caption{List of thermo-optic phase shifters investigated in this work} 
\begin{tabular}{c | c | c | c | c | c}
\hline
 Device & \begin{tabular}{@{}c} Trench \& \\undercut \end{tabular}&  \begin{tabular}{@{}c} Waveguide \\ passes \end{tabular} & \begin{tabular}{@{}c} Waveguide \\ length under \\ TiN (mm) \end{tabular} & \begin{tabular}{@{}c} TiN width \\ ($\mu$m) \end{tabular}& \begin{tabular}{@{}c} Suspended region \\  area ($\mu$m $\times \mu$m) \end{tabular} \\ [0.5ex] 
 \hline
 PS1 & Yes & 3 & 1.5 & 4.2 & $8.2 \times 500$ \\ 
 \hline
 PS2 & No & 3 & 1.5 & 4.2 & N.A. \\ 
 \hline
 PS3 & Yes & 7 & 6.9 & 10 & $15.4 \times 984$ \\ 
 \hline
\end{tabular}
\end{table}

Three phase shifter designs were investigated, and the parameters of each design are summarized in Table I. PS1 is the design described in this section and shown in Fig. \ref{fig:fig1}.  PS2 was nominally identical PS1 except for the lack of deep trenches and undercut etching. PS3 was investigated to explore the trade-off between the device insertion loss and tuning efficiency by increasing the length of the SiN waveguides under the TiN header; the 7 SiN waveguides under the heater had widths alternating among 600, 650, and 550 nm.

\begin{figure}
\centering
     \begin{subfigure}[b]{1\textwidth}
         \centering
         \includegraphics[width=0.8\textwidth]{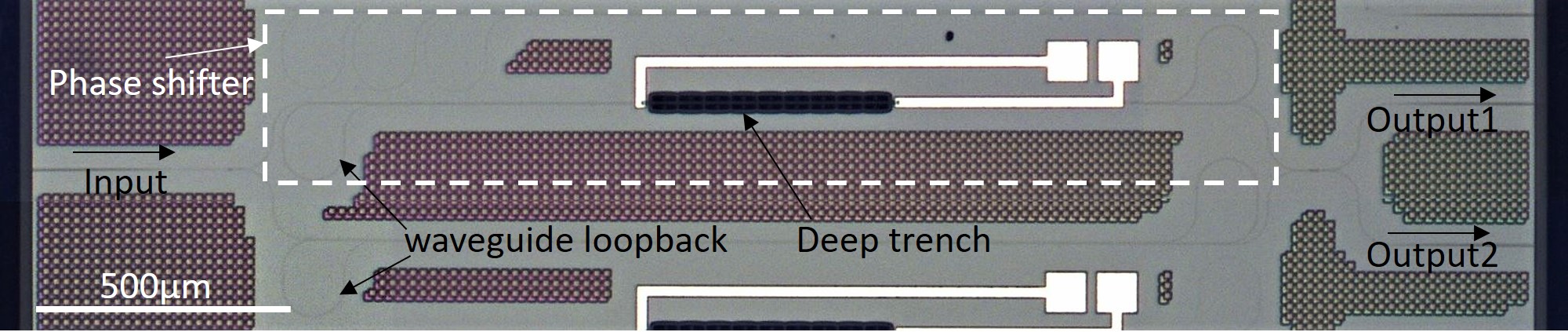}
         \subcaption{}
         \label{figure2(a)}
     \end{subfigure}
      \begin{subfigure}[b]{1\textwidth}
         \centering
         \includegraphics[width=0.8\textwidth]{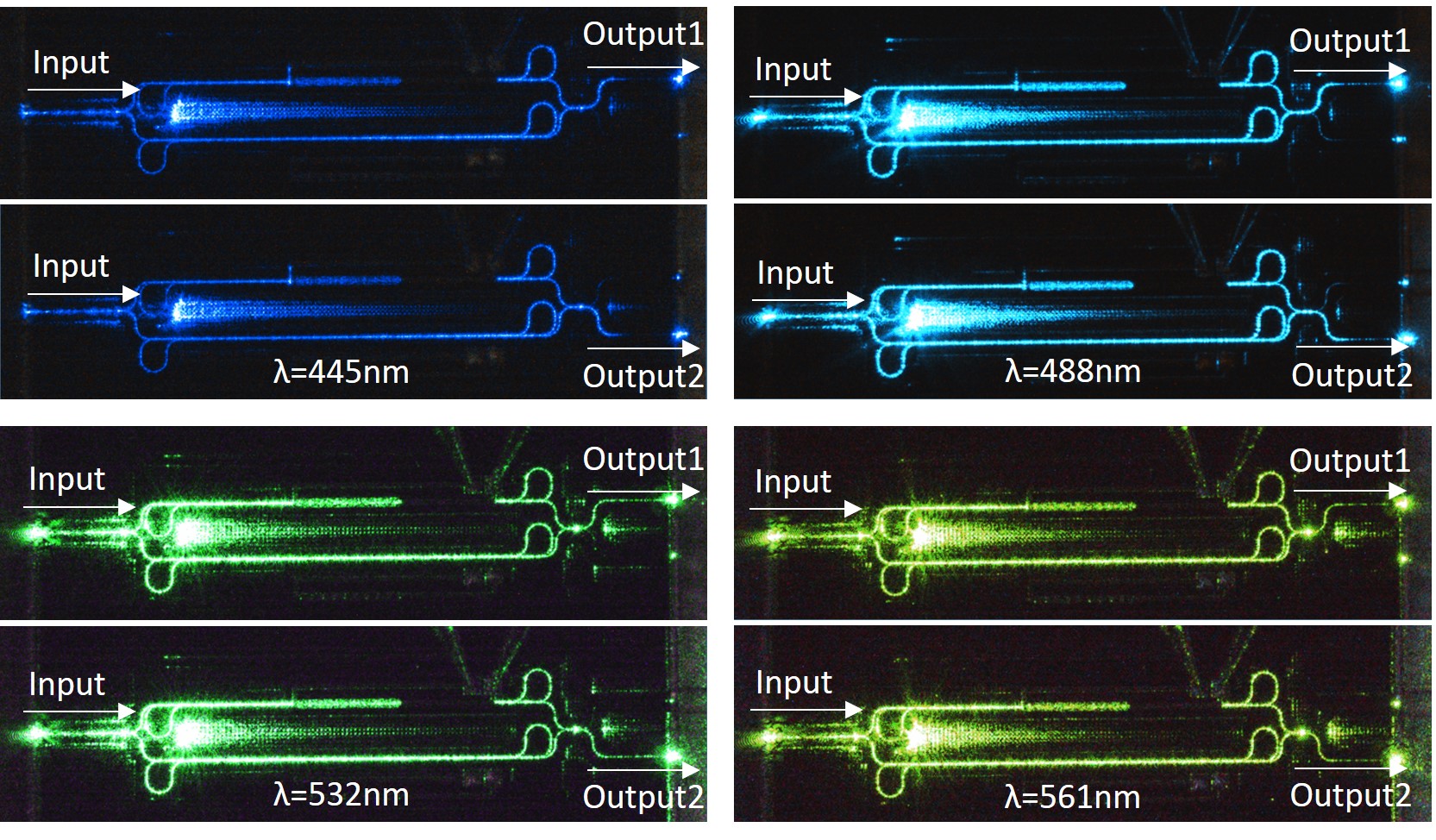}
         \subcaption{}
         \label{figure2(b)}
     \end{subfigure}
\hfill
      \begin{subfigure}[b]{1\textwidth}
         \centering
         \includegraphics[width=0.8\textwidth]{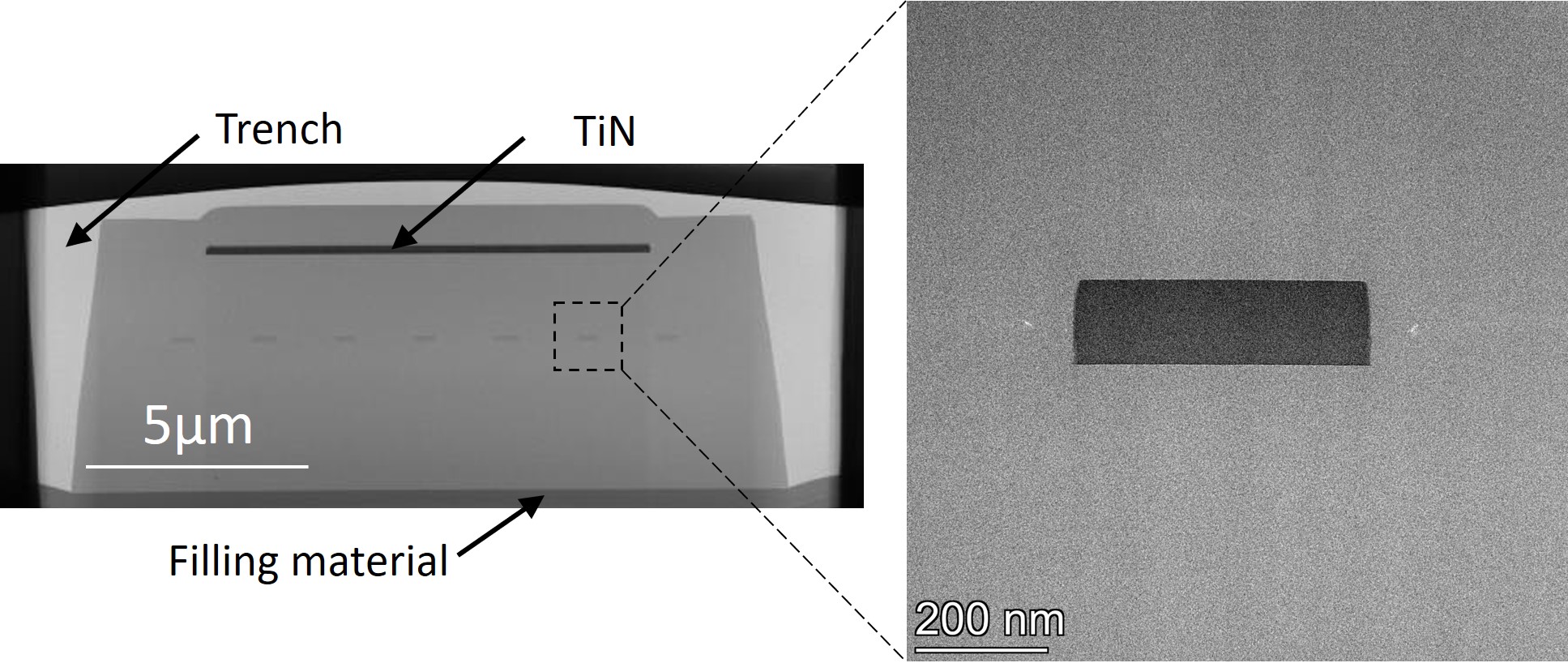}
         \subcaption{}
         \label{figure2(c)}
     \end{subfigure}
\caption{(a) Optical micrograph of PS1 integrated into a MZI. (b) Optical micrographs of the MZI with optical inputs at various  wavelengths; at each wavelength, the phase shifter was driven to demonstrate switching between the MZI output ports. (c) Cross-section transmission electron micrograph (XTEM) of the suspended region of PS3 showing 7 SiN waveguides under the TiN heater; (inset) magnified view of one of the SiN waveguides. The filling material required for the XTEM is delineated.}
\end{figure}

\section{Device fabrication and measurements}
\label{sec:measurements}

The thermo-optic phase shifters were fabricated on 200-mm diameter Si wafers at Advanced Micro Foundry (AMF) as part of our visible-light integrated photonics platform. The fabrication process, which includes multiple steps to define other devices in the platform, began with Si mesa patterning and doping to define Si photodiodes \cite{yiding2021pd}. Next, PECVD SiO$_2$ deposition and planarization was performed to define the bottom cladding of the waveguides. SiN waveguides (150 nm nominal thickness) were fabricated by SiN PECVD, deep ultraviolet (DUV) lithography, and reactive ion etching (RIE). Additional SiO$_2$ and SiN deposition and patterning steps were performed to define a second SiN waveguide layer (75 nm nominal thickness) for bi-layer edge couplers \cite{yiding2021ec}. Chemical mechanical polishing (CMP) was used for layer planarization. Following deposition of SiO$_2$ top cladding for the waveguides, TiN heaters were defined, followed by 2 metal wiring layers (M1 and M2), vias (M1-M2, TiN-M2, Si photodiode - M1), and oxide openings for bond pads; similar to conventional Si photonic platforms for infared wavelengths \cite{sacher2018monolithically}. Finally, deep trench and undercut etching were performed to define facets for edge couplers and the suspended SiO$_2$ bridges of the thermo-optic phase shifters.    

Figure 2(a) shows an optical micrograph of PS1 integrated into a MZI test structure. The input and output fiber-to-chip tapered edge couplers are labeled in the figure and have a width of 5.2 $\mu$m at the facets. Optical micrographs of the MZI tested at blue, green, and yellow wavelengths are shown in Fig. 2(b); switching between the two outputs of the MZI by driving the phase shifter is demonstrated.

Figure 2(c) shows a cross-section transmission electron micrograph (XTEM) of PS3; the cross-section cut through the suspended SiO$_2$ bridge and has the same orientation as the schematic in Fig. \ref{fig:fig1}(b). The device measurements reported in this work focus on one wafer from the fabricated wafer lot, while the XTEM is from a second wafer. The only fabrication difference between these wafers was the SiO$_2$ bottom cladding thickness of the SiN waveguides; 3 $\mu$m for the first wafer and 3.25 $\mu$m for the second. Investigation of one die from the second wafer yielded a similar PS1 performance to the first wafer. From the XTEM, the average separation between the SiN waveguide and the TiN heater was about 1.86 $\mu$m, and the suspended region had a trapezoidal cross-section. The average separation between the edge SiN waveguide to the left and right edges of the suspended region was about 2 $\mu$m, as designed. The average SiN waveguide width was about 100 nm narrower compared to the designed value, and correspondingly, the average SiN waveguide gap was about 100 nm larger than designed. The average SiN waveguide thickness in the XTEM was about 130 nm.

\begin{figure}
\centering
    \subfloat[\centering ]{{\includegraphics[width=0.4\textwidth]{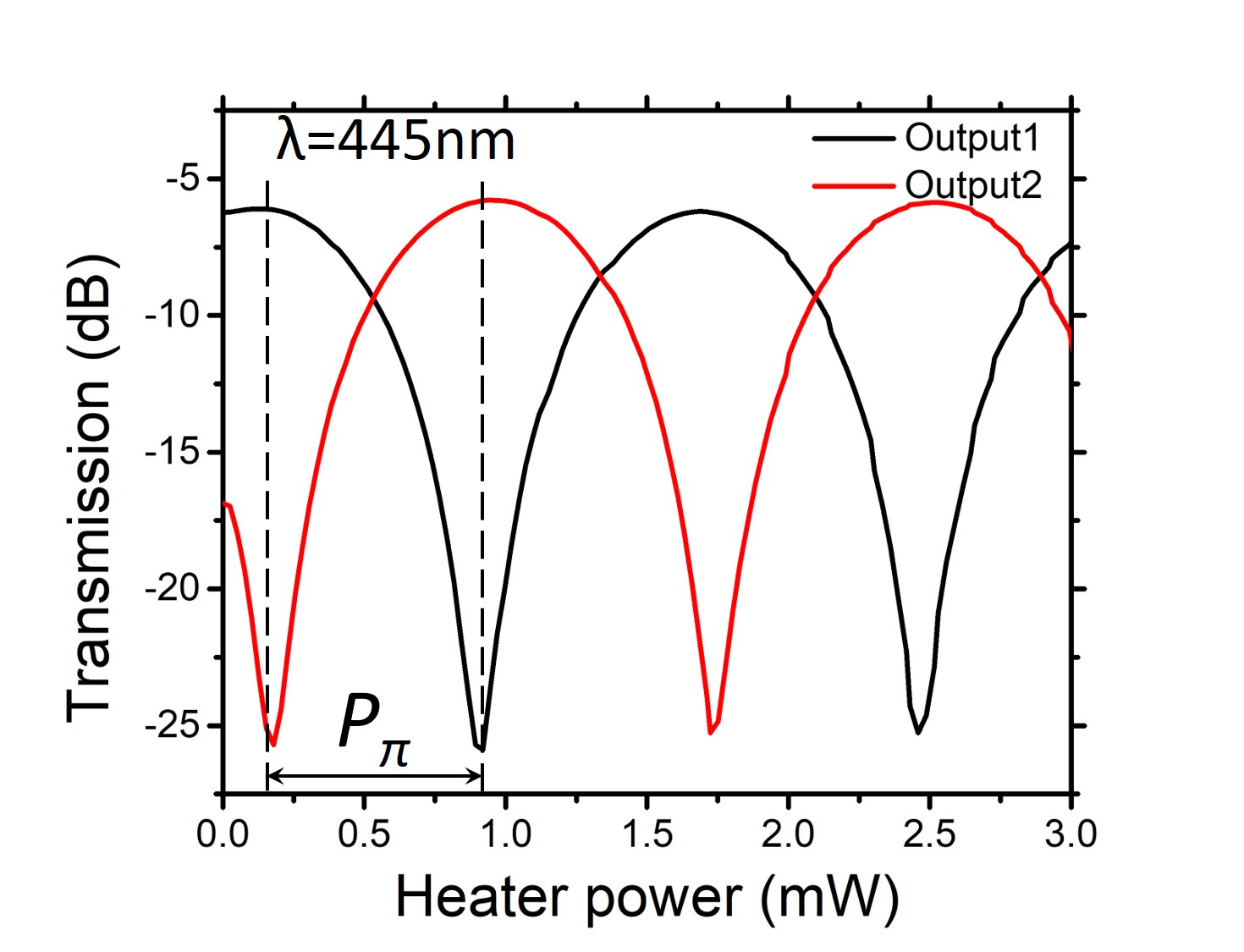} }}%
    \subfloat[\centering ]{{\includegraphics[width=0.4\textwidth]{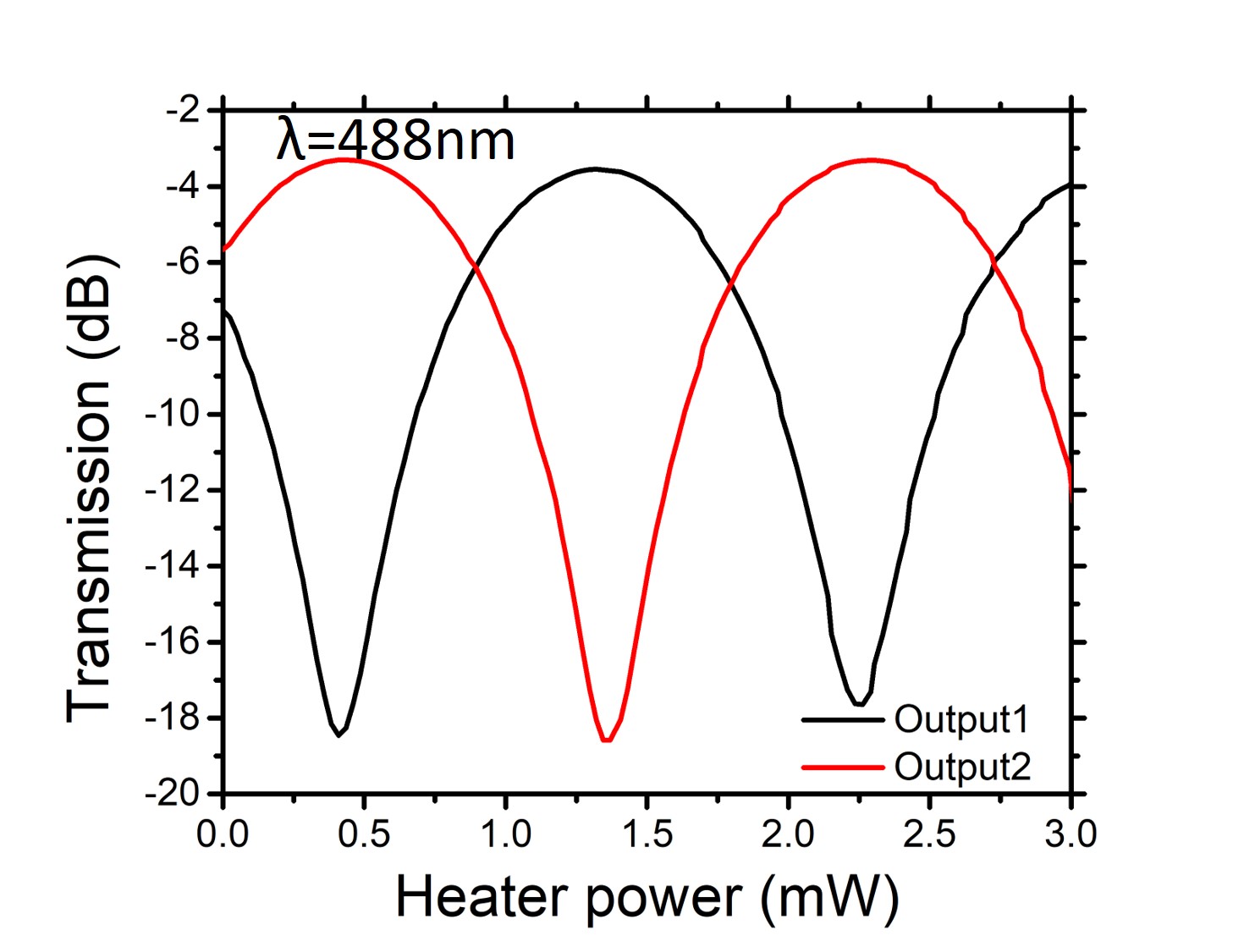} }}%
    \qquad
    \subfloat[\centering ]{{\includegraphics[width=0.4\textwidth]{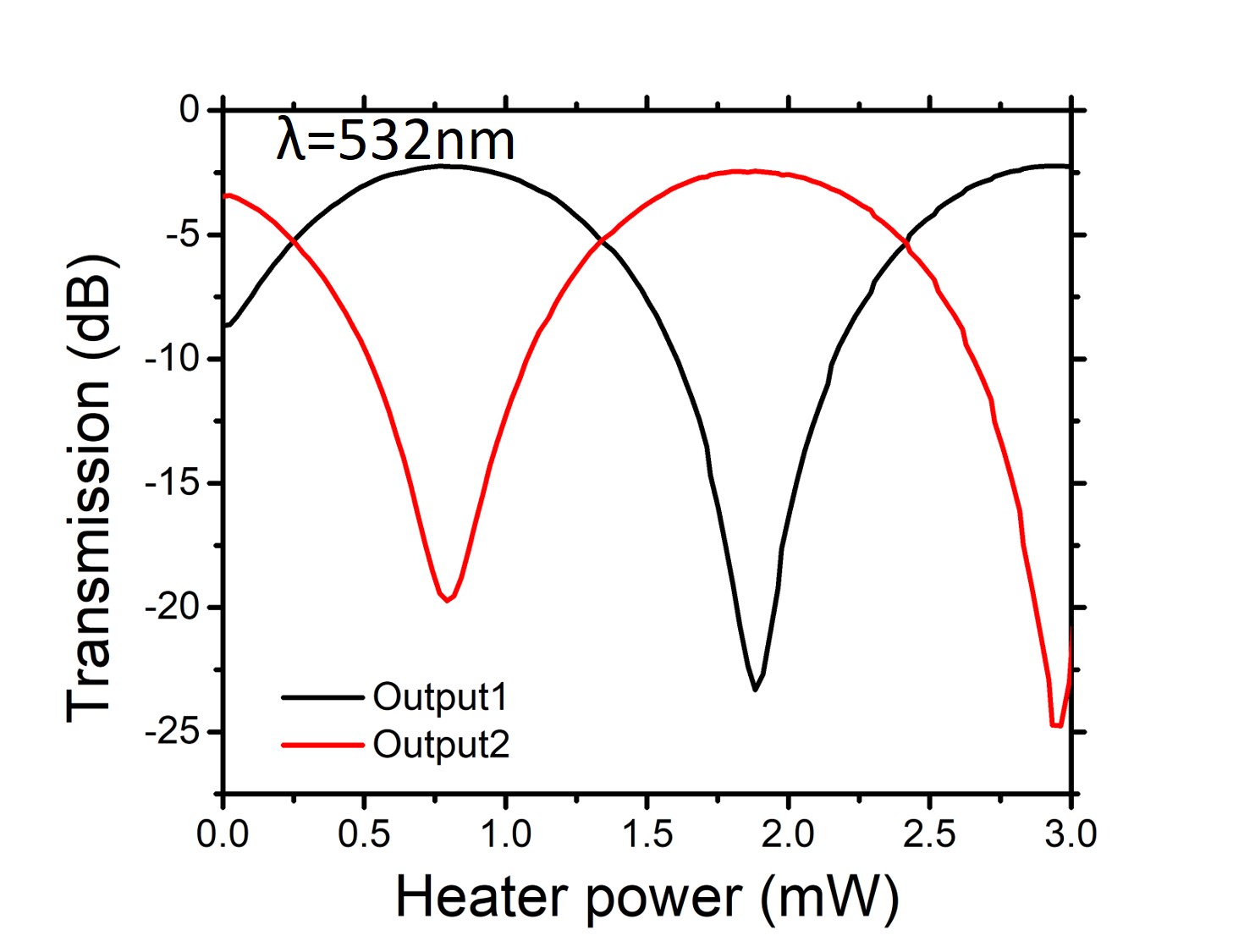} }}%
    \subfloat[\centering ]{{\includegraphics[width=0.4\textwidth]{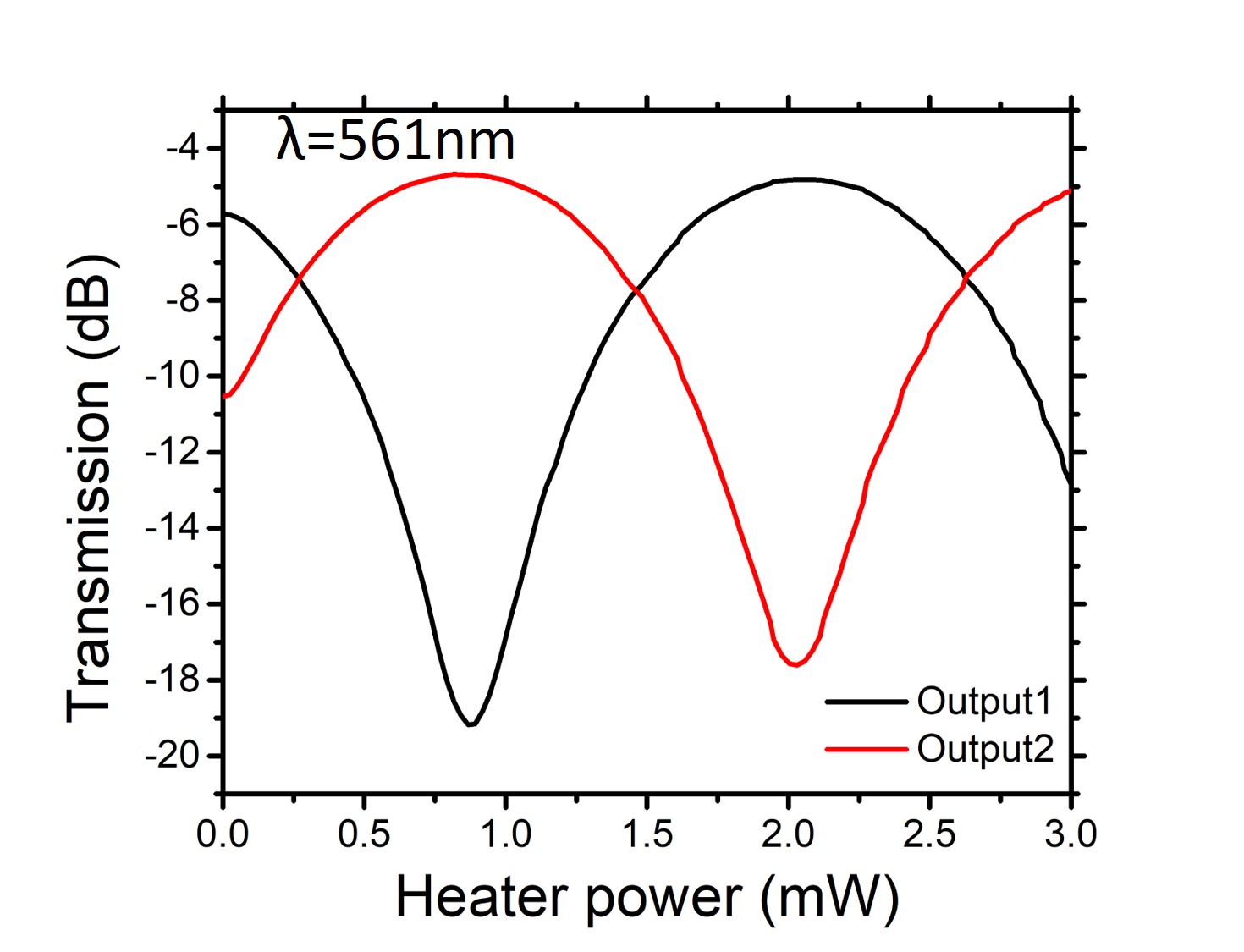} }}%
\caption{Measured optical transmission vs. heater power dissipation for the PS1 MZI [as shown in Fig. 2(a)] at wavelengths of (a) 445 nm, (b) 488 nm, (c) 532 nm, and (d) 561 nm. Traces are shown for each MZI output ("Output1" and "Output2"), and $P_{\pi}$ is defined in (a).}
\end{figure}

Device characterization was performed using Coherent OBIS LX  (445 and 488 nm) and Sapphire FP (532 and 561 nm) lasers and a Newport 818-SL/DB photodetector connected to a Newport 2936-R optical power meter. Cleaved single-mode fibers (Nufern S405-XP) coupled light on-off the chips via the on-chip edge couplers, as in \cite{sacher2019visible}. The input fiber had an inline fiber polarization controller to set the input polarization. DC drive voltages were applied to the devices from a sourcemeter (Keysight B2912A) via tungsten electrical probes contacting the on-chip pads, and the measured PS1 TiN heater resistance was about 1050 $\Omega$. 

The phase shifters were integrated into MZI test structures to convert the applied phase shift into transmission changes at the MZI outputs. The measured PS1 MZI transmission as a function of heater power is shown in Fig. 3 at $\lambda = $ 445, 488, 532, and 561 nm. The power required to drive the MZI between the cross and thru states is $P_\pi$ of the phase shifter. The transmissions were normalized to those of a reference waveguide with the same edge coupler design as the MZI. The peaks and nulls of the transmission traces also indicate the insertion loss (IL) and extinction ratio (ER) of the MZI switch. The IL, ER, and $P_\pi$ were calculated from the first half-period of the tuning. Slight differences in the measurements from the two output ports were observed; likely due to waveguide loss variations between the MZI output waveguides, amplitude and phase errors of the $2\times 2$ MMI, and measurement error. Due to fiber alignment error and the waveguide loss of the reference waveguide being subtracted from the MZI loss during normalization, we estimate the IL measurements to be accurate to $\approx \pm 1.5$ dB.

The measured results of PS1 are summarized in Table II. The phase shifter exhibited a low power consumption with $P_\pi$ measurements ranging from 0.79 - 1.22 mW and increasing with wavelength. The simulated $P_\pi$ values in Table II are in close agreement with the measurements. The increase of $P_\pi$ with wavelength is a result of the reduced optical confinement in the SiN at longer wavelengths and the lower thermo-optic coefficient of SiO$_2$ relative to SiN. The ER was $> 13$ dB in all cases, indicating that the deep trenches in the phase shifter did not cause significant excess losses that would otherwise imbalance the MZI. Some dependence of the ER on the heater power was observed.

The IL of the PS1 MZI is a combination of the MMI losses and the waveguide losses in the MZI arms. Cutback measurements of 550 nm wide waveguides on one die indicated the waveguide losses were about 8.2, 7.0, 5.5, and 5.8 dB/cm at $\lambda = $ 445, 488, 532, and 561 nm, respectively. Simulations indicated minimum excess losses of 0.2 - 0.3 dB per MMI at green wavelengths and larger losses expected at $\lambda = $ 445 and 561 nm.

\begin{table}[t!]
\centering
\caption{Summary of the measurements of the primary phase shifter (PS1) integrated into a MZI} 
\begin{tabular}{c | c | c | c | c | c | c | c}
\hline
 \begin{tabular}{@{}c} $\lambda$ \\ (nm) \end{tabular}& \multicolumn{2}{c|} {\begin{tabular}{@{}c} IL \\(dB) \end{tabular}} &  \multicolumn{2}{c|} {\begin{tabular}{@{}c} ER \\ (dB) \end{tabular}} & \multicolumn{3}{c} {\begin{tabular}{@{}c} $P_\pi$ \\ (mW) \end{tabular}} \\ [0.5ex] 
 \hline
   & Output1 & Output2 & Output1 & Output2 & Output1 & Output2 & Simulation \\ 
 \hline
 445 & 6.1 & 6.5 & 19.8 & 19.0 & 0.79 & 0.82 & 0.91 \\ 
 \hline
 488 & 3.6 & 3.9 & 14.9 & 15.3 & 0.91 & 0.94 & 1.03 \\ 
 \hline
 532 & 2.5 & 2.4 & 21.0 & 17.3 & 1.09 & 1.09 & 1.17 \\ 
 \hline
 561 & 4.8 & 4.9 & 14.4 & 13.2 & 1.22 & 1.16 & 1.27 \\ 
 \hline
\end{tabular}
\end{table}

The rise and fall times of the PS1 MZI are presented in Fig. 4. A square wave voltage signal from a function generator (Tektronix AFG 31000) was applied to the device. The output optical signal was detected and read out on an oscilloscope (New Focus 2032, Keysight DSOX4054A). For Fig. 4(a), the `1' level average voltage was about 0.500 V, and the  `0' level average voltage was about -0.239 V; for Fig. 4(b), the measured `1' level average voltage was about 0.503 V, the measured `0' level average voltage was about -0.239 V. The estimated $10-90$\% rise and fall times were about 570 $\mu$s and 590 $\mu$s, respectively.

\begin{figure}[t!]
\centering
    \subfloat[\centering ]{{\includegraphics[width=0.4\textwidth]{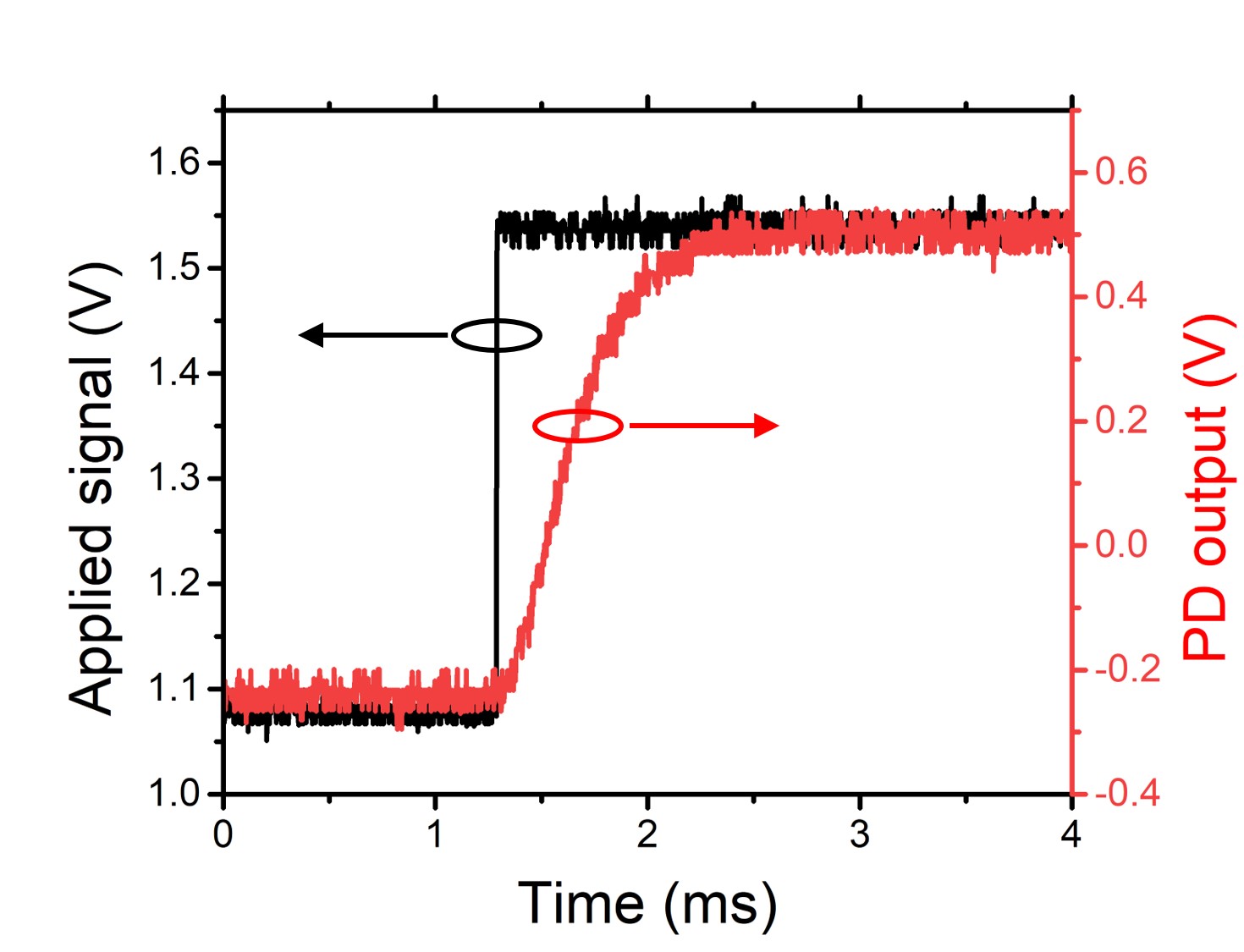} }}%
    \subfloat[\centering ]{{\includegraphics[width=0.4\textwidth]{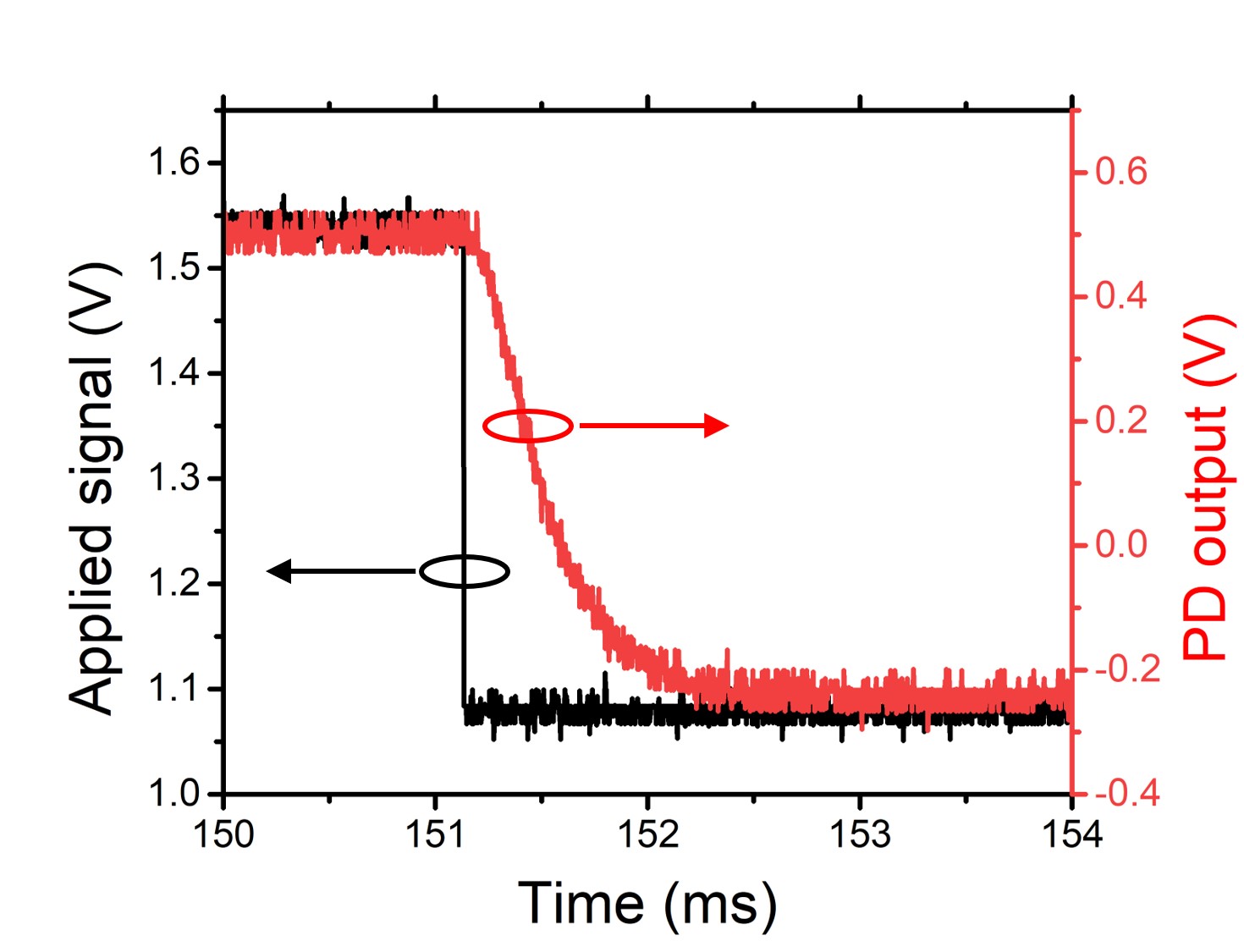} }}%
    \label{figure4}%
\caption{Step response measurements of the PS1 MZI showing the (a) rise and (b) fall time. Each plot shows the applied voltage to the heater vs. time and the detected optical signal.}
\end{figure}

\begin{figure}
\centering
    \subfloat[\centering ]{{\includegraphics[width=0.4\textwidth]{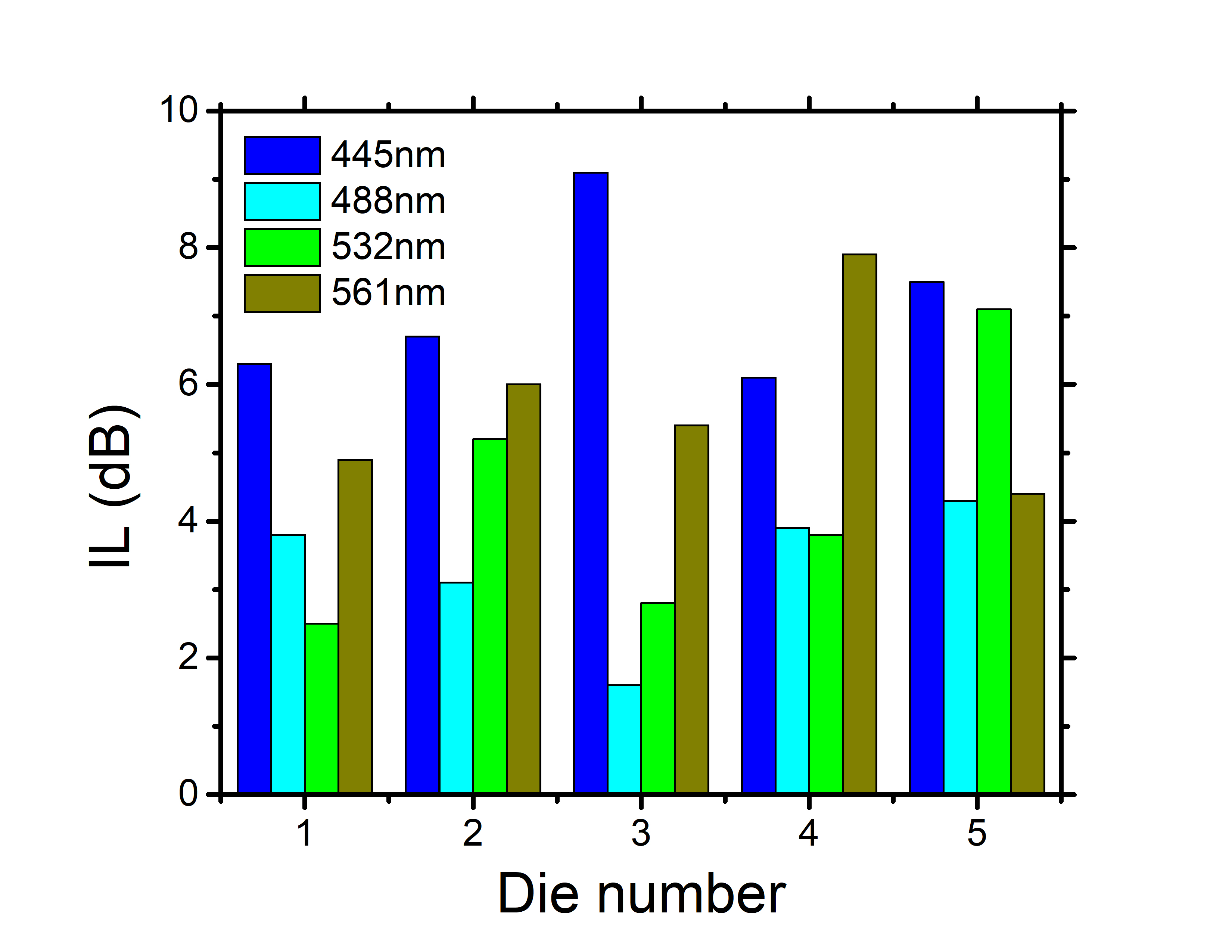} }}%
    \subfloat[\centering ]{{\includegraphics[width=0.4\textwidth]{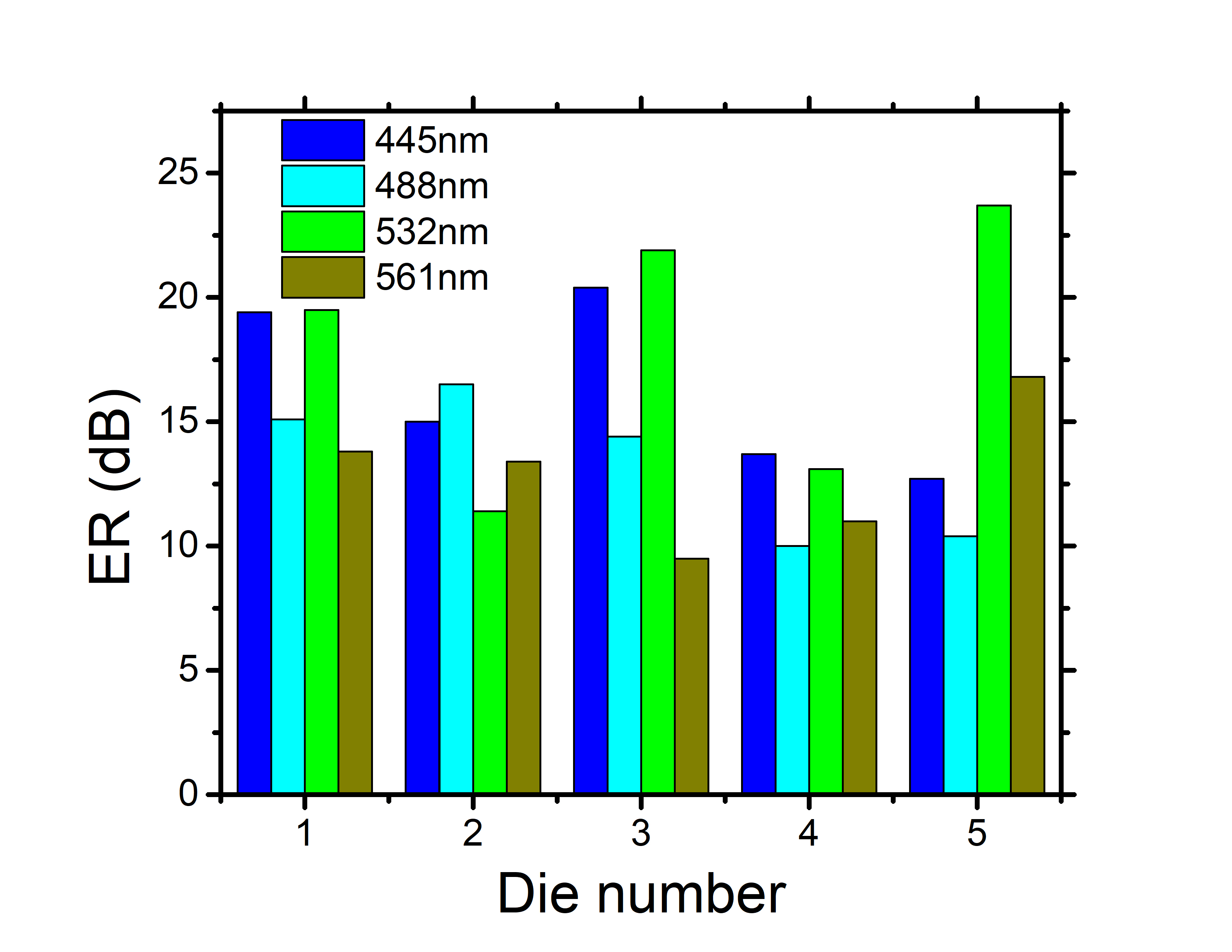} }} \\
    \subfloat[\centering ]{{\includegraphics[width=0.4\textwidth]{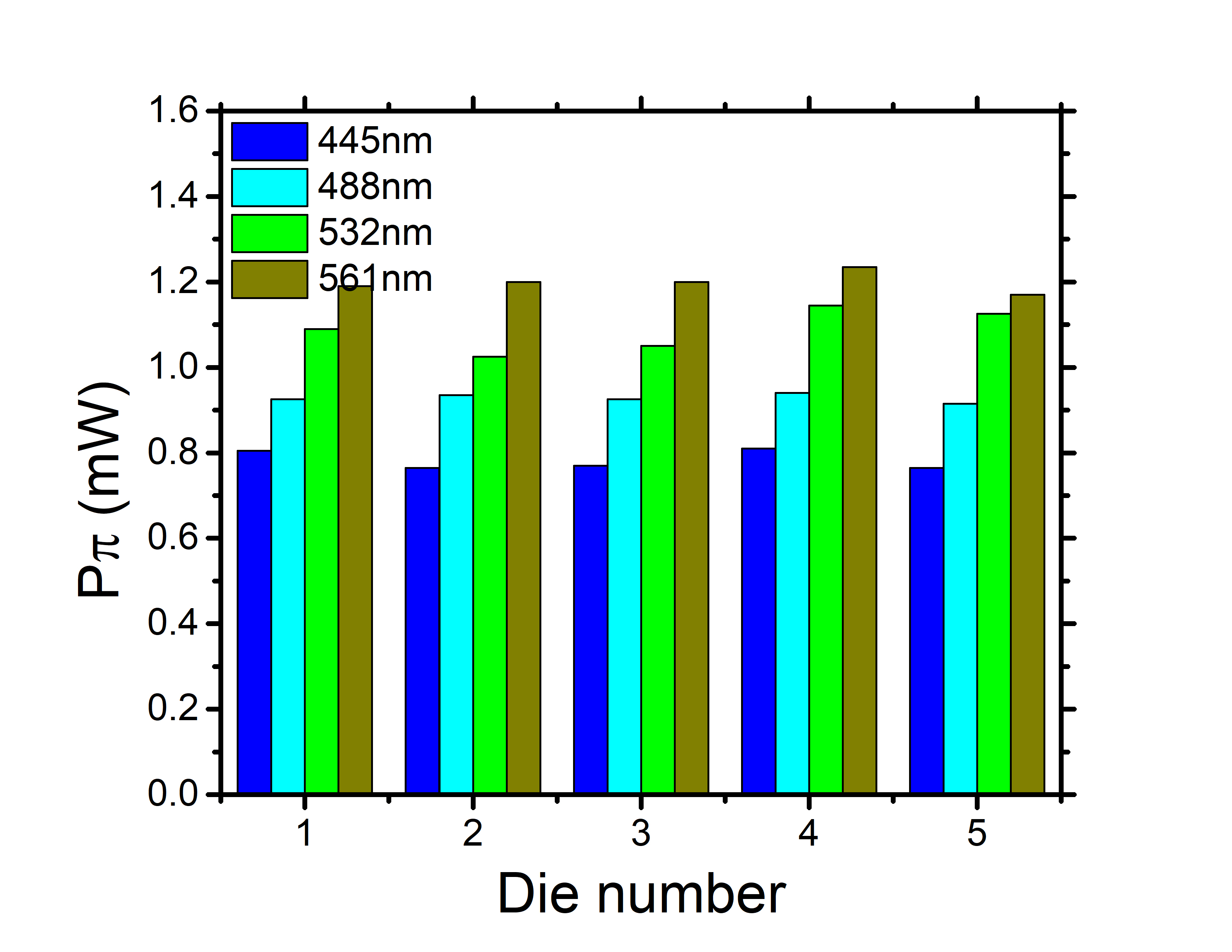} }}%
    \subfloat[\centering ]{{\includegraphics[width=0.25\textwidth]{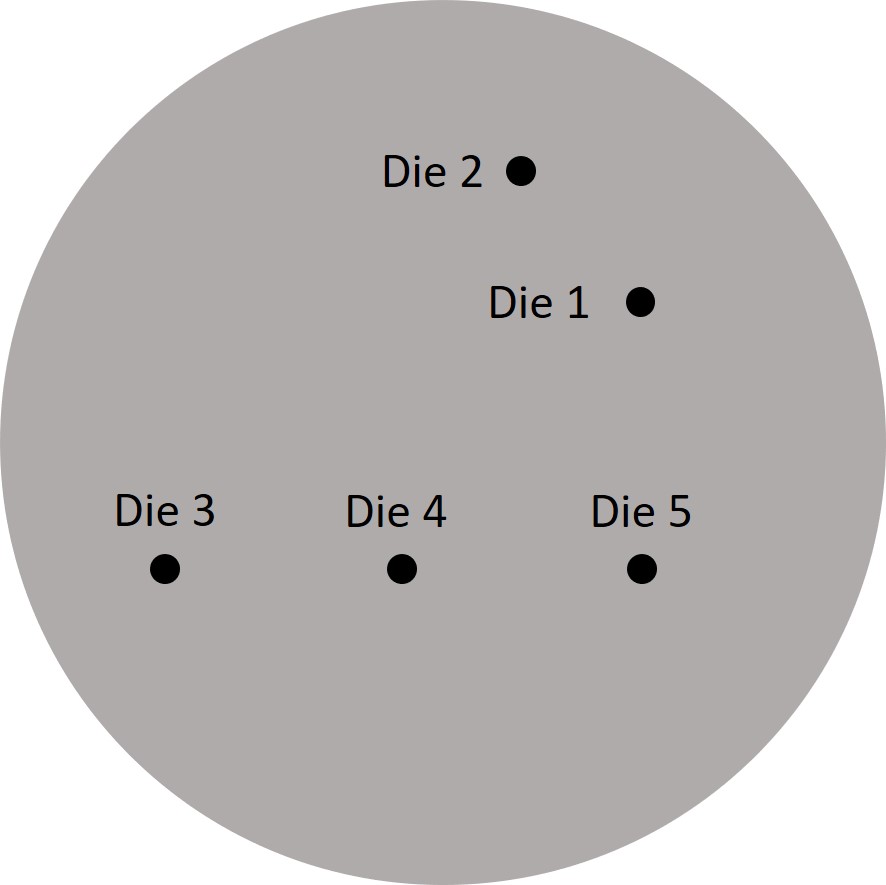} }}%
    \label{figure5}%
\caption{Measured (a) IL, (b) ER, and (c) $P_\pi$ of the PS1 MZI from 5 different dies across the wafer. The measurements in Fig. 3 are from Die 1. (d) The locations of different dies measured.}
\end{figure}

To check the repeatability of the results, we measured the tuning curves of the PS1 MZI on five different dies from the same wafer, and the results are summarized in Fig. 5(a)-(c), with the die locations shown in Fig. 5(d). The values in Fig. 5 were calculated from the average of the first half-period tuning of the MZI output ports 1 and 2. The average $P_\pi$ of PS1 across the dies was 0.78 mW, 0.93 mW, 1.09 mW, and 1.20 mW at $\lambda = $ 445, 488, 532, and 561 nm, respectively, and the variation in $P_\pi$ across the dies for each wavelength was $\leq 0.1$ mW. The average ER was 17.3 dB, 14.0 dB, 20.1 dB, and 13.6 dB at the same wavelengths, and the ER was $\geq 9.5$ dB in all cases. The average IL across the dies was 7.3 dB, 3.4 dB, 4.6 dB, and 5.9 dB at $\lambda = $ 445, 488, 532, and 561 nm, respectively; correspondingly, the maximum/minimum IL at each wavelength was 9.1/6.1 dB, 4.3/1.6 dB, 7.1/2.5 dB, 7.9/4.4 dB. 

Contributions to the observed IL and its variation include measurement uncertainties (due to fiber alignment and laser power fluctuations), material and device loss across the wafer, and on-chip Fabry-Perot oscillations. The measurements were performed using 4 separate single-wavelength lasers with different spectral/power stabilities; the 532 nm laser had the largest observed fluctuations. Power fluctuations resulted in uncertainties in transmission measurements and polarization settings, both contributing to IL uncertainty; wavelength fluctuations affected the measured ER. 

Due to the high mode confinement in the waveguides and the relatively small waveguide dimensions required for single- or few-mode operation in the visible spectrum, the waveguide losses are more sensitive to sidewall roughness scattering (compared to standard silicon photonics at infrared wavelengths). Similarly, the MMI losses are more sensitive to waveguide dimension errors. The device performance can likely be more consistent with optimization of the SiN etch process.

\section{Discussion}
\label{sec:discussion}
To investigate the influence of design parameters, we compared the performance of PS1 against PS2 and PS3 (Table I). PS2, which did not have the deep trenches and undercut, had similar measured IL and ER as PS1 (comparing the MZI measurements); however, as expected, it had lower efficiency. The $P_\pi$ of PS2 was 15.9, 18.4, 22.2, and 23.0 mW at $\lambda = 445$, 488, 532, and 561 nm respectively, about $20 \times$ higher than PS1. The benefit was a much shorter $10-90$\% rise(fall) time of 34.5(33.5) $\mu$s.

PS3 had 7 waveguide passes in the suspended heater (compared to 3 for PS1) with the aim of lower $P_{\pi}$ (at the expense of higher IL). PS3 simulations and measurements are detailed in the Appendix. The measured $P_{\pi}$ of PS3 was 0.65 mW at $\lambda = 445$ nm, about 81\% of PS1. The small reduction in $P_{\pi}$ was accompanied by a large increase in insertion loss due to the increased waveguide length. As explained in the Appendix, the 15.4 $\mu$m suspended region width (1.9$\times$ wider than that of PS1) resulted in an incomplete undercut in between the SiO$_{2}$ anchors, and consequently, a reduced thermo-optic power efficiency. 

Reductions in $P_{\pi}$ of PS1 may be possible through relatively simple design optimizations. Reducing the number of SiO$_{2}$ anchors supporting the suspended region is expected to substantially increase the power efficiency \cite{lu2015michelson}, possibly at the expense of reduced mechanical robustness. In addition, according to the simulations in Fig. \ref{fig:fig1}(c), our chosen waveguide gap of 1.2 $\mu$m could be reduced to 0.6 $\mu$m with a maximum crosstalk $< -20$ dB at $\lambda = 561$ nm, and the corresponding reduction in the suspended region volume is expected to increase the phase shifter power efficiency.

The insertion loss measurements in Section \ref{sec:measurements} include both the phase shifter and the MMIs of the MZI. Separate test structures to measure the loss of the MMIs were not available, and thus, it was not possible to directly measure the phase shifter loss from the MZI. Instead, as mentioned in Section \ref{sec:measurements}, we measured the waveguide loss of a 550 nm waveguide (the smallest of the 3 waveguide pass widths in the heater); the loss ranged from about 8 - 5.5 dB/cm from $\lambda =$ 445 - 561 nm, the total phase shifter waveguide length was 4.259 mm, and the resulting estimated phase shifter loss is 3.4 - 2.3 dB. The insertion loss of the phase shifter can be reduced by minimizing the waveguide length that is not in the heated suspended region. In this initial demonstration, a significant length of unnecessary routing waveguides was present in PS1. This is evident in Fig. \ref{fig:fig1}(a); the waveguide bends for the loopbacks extend 521 $\mu$m beyond the suspended region on each side. By removing this unnecessary waveguide length and the waveguide length used for routing from the MZI MMIs to and from the phase shifter, the minimum PS1 waveguide length is 3.335 mm; the estimated loss of this phase shifter is 2.7 - 1.9 dB from $\lambda =$ 445 - 561 nm. 

The phase shifter loss can also be reduced by minimizing the waveguide scattering loss. With further optimization of the fabrication process, we expect the waveguide loss to be reduced to, at most, our previously demonstrated passive SiN waveguide platform losses (4 - 2 dB/cm for similar waveguide dimensions from $\lambda =$ 445 - 561 nm) \cite{sacher2019visible}. With these reduced waveguide losses and the length optimization mentioned above, we expect the phase shifter loss could be about 1.3 - 0.7 dB from $\lambda =$ 445 - 561 nm.

\begin{figure}
\centering
    \subfloat[\centering ]{{\includegraphics[width=0.4\textwidth]{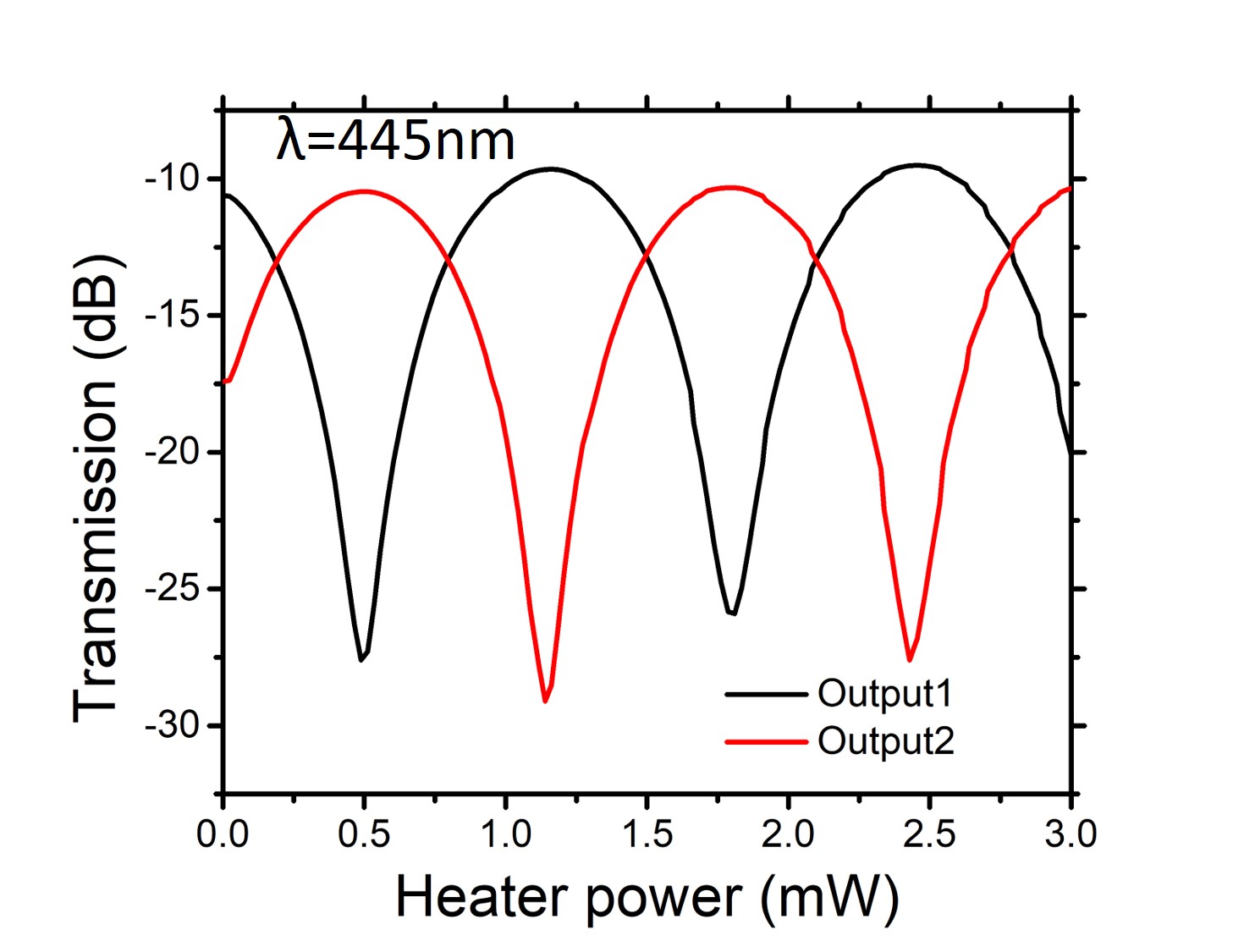} }}%
    \subfloat[\centering ]{{\includegraphics[width=0.4\textwidth]{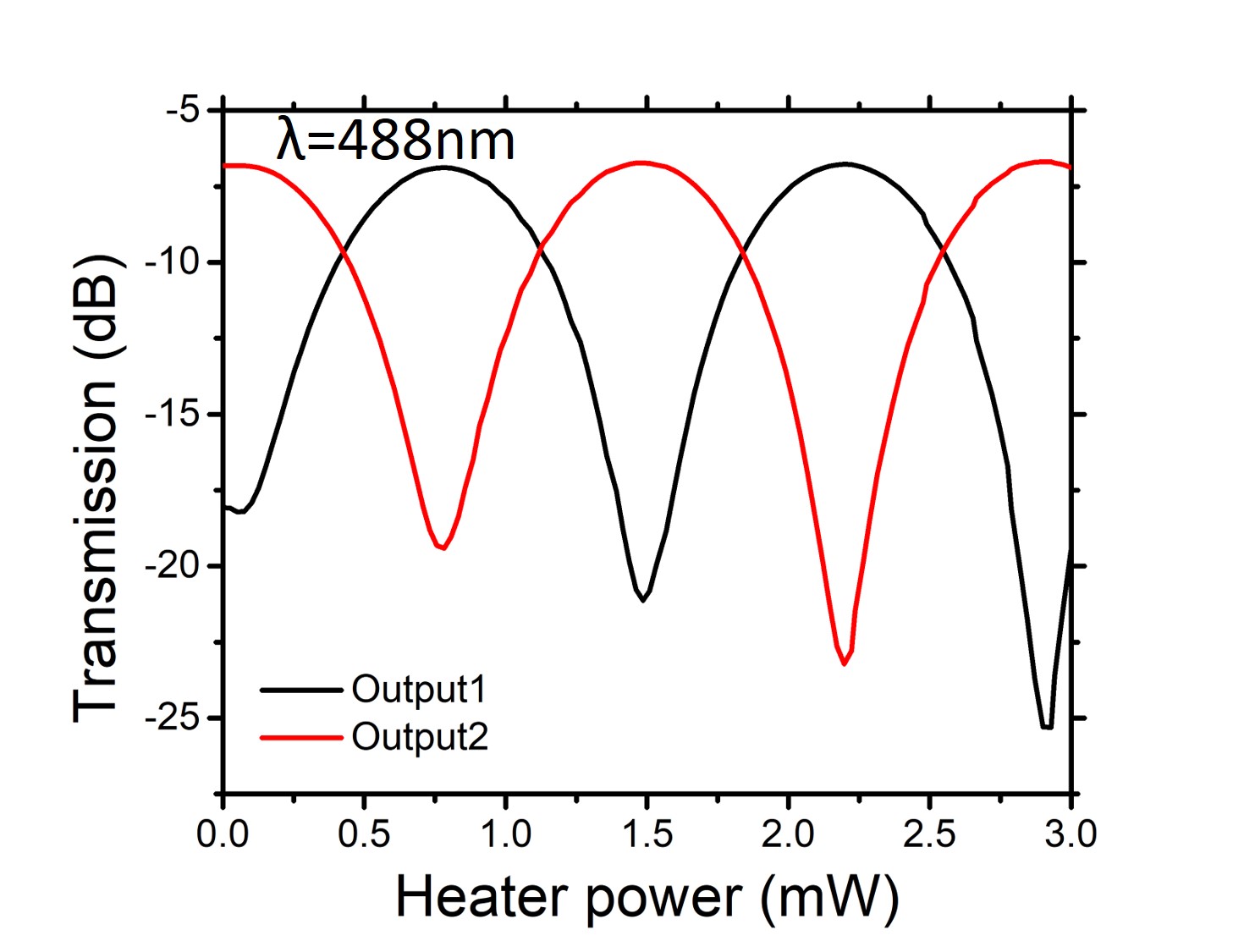} }}%
    \qquad
    \subfloat[\centering ]{{\includegraphics[width=0.4\textwidth]{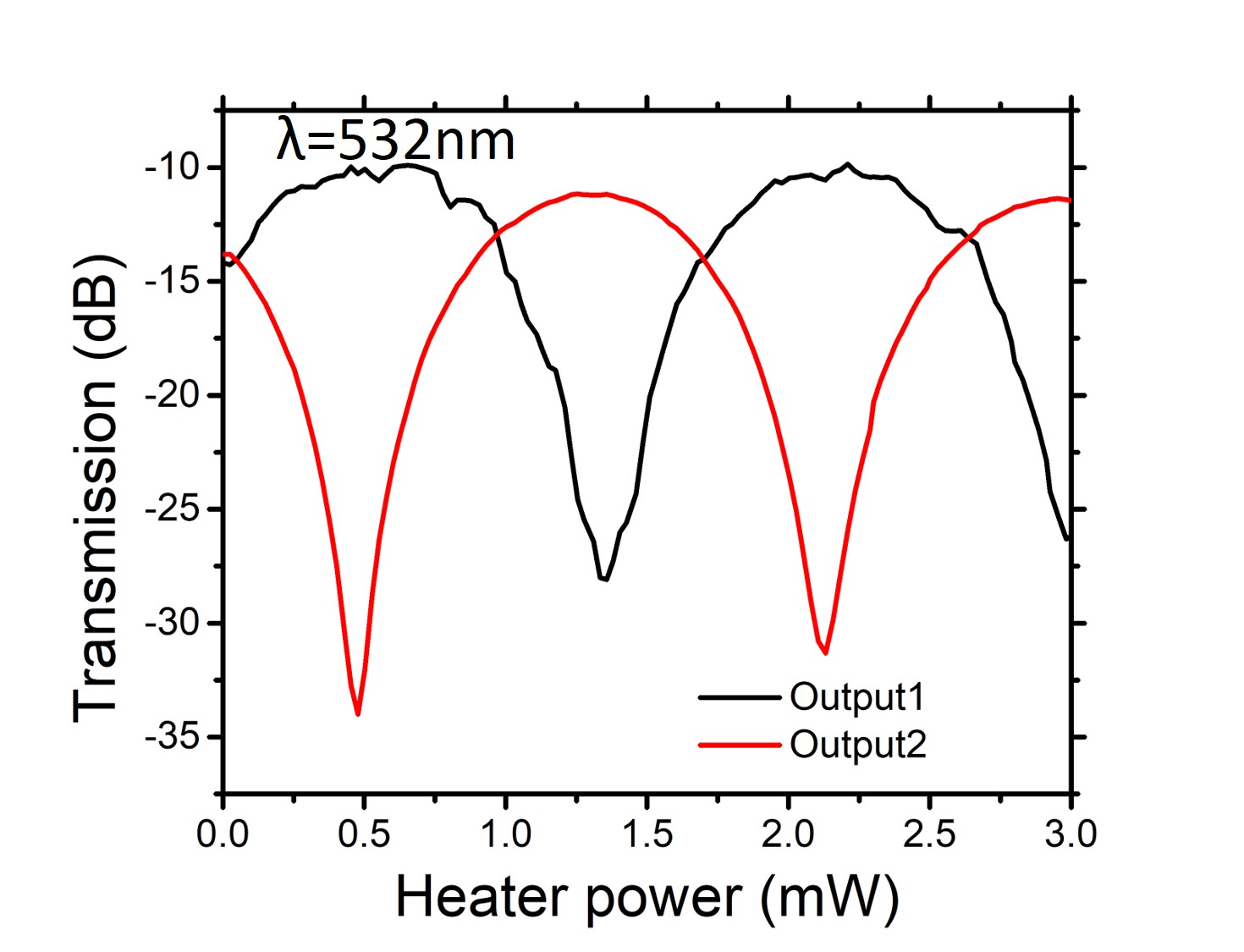} }}%
    \subfloat[\centering ]{{\includegraphics[width=0.4\textwidth]{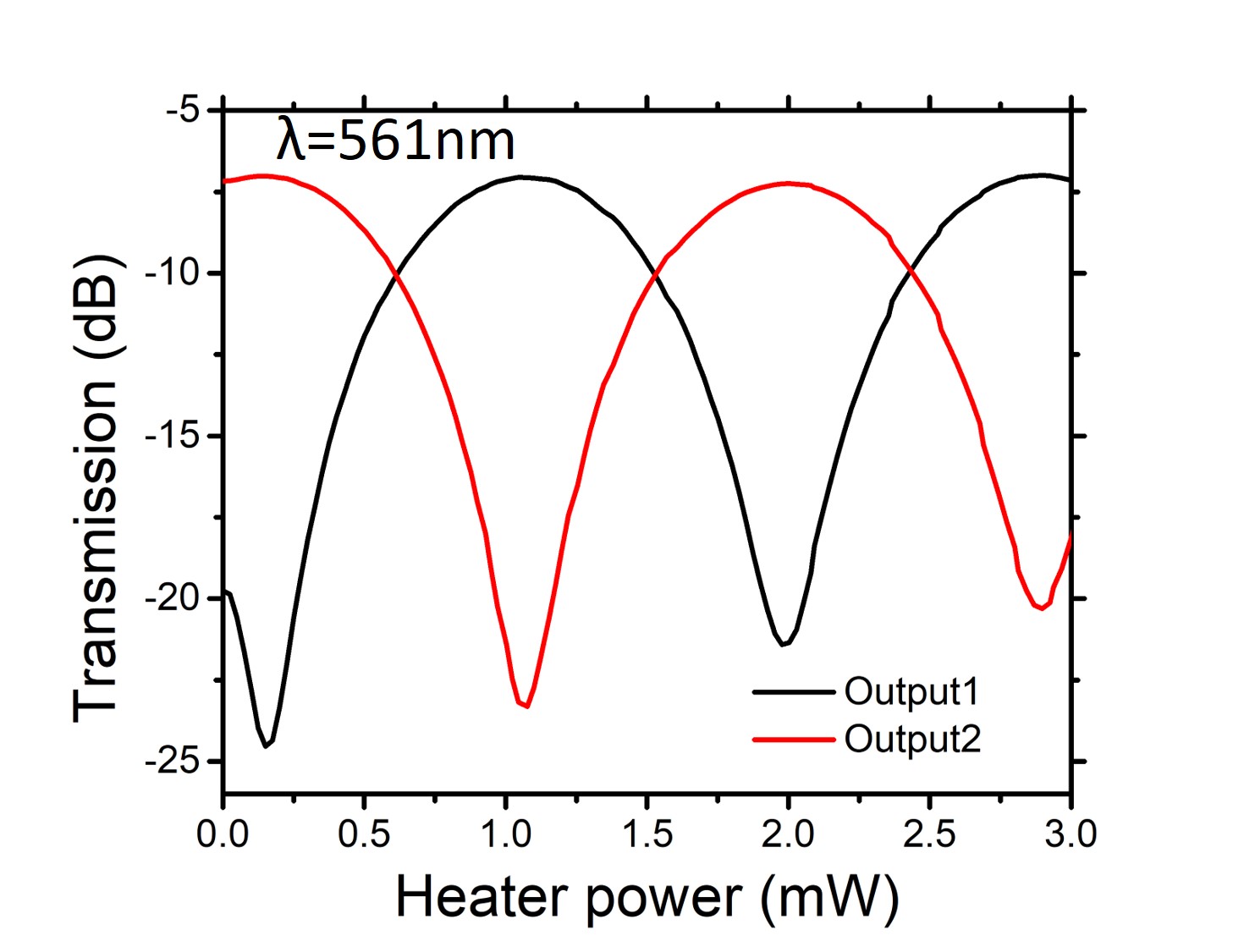} }}%
\caption{Measured transmission vs. heater power dissipation for the PS3 phase shifter integrated into a MZI at wavelengths of (a) 445 nm, (b) 488 nm, (c) 532 nm, and (d) 561 nm. Traces are shown for each MZI output ("Output1" and "Output2").}
\end{figure}

\section{Conclusion}
\label{sec:conclusion}
In summary, we have demonstrated power-efficient SiN thermo-optic phase shifters for the visible spectrum. The phase shifters had an ultra-low average $P_{\pi}$ of 0.78 mW at $\lambda = 445$ nm. The phase shifters were integrated into MZI switch test structures, and the measured 10\% to 90\% rise and fall times were about 570 $\mu$s and 590 $\mu$s, respectively. The devices operated at blue, green, and yellow wavelengths; the optical bandwidth was limited by the MMI couplers of the MZI. The phase shifters were fabricated on 200-mm diameter Si wafers as part of our visible-light integrated photonics platform. These demonstrated devices are monolithically integrated into silicon photonic circuits in contrast to phase shifters based on liquid crystals and electro-optic materials. They provide new possibilities to implement phase shifters in large-scale active photonic circuits in the visible spectrum.

\begin{table}
\centering
\caption{Summary of the measurements of PS3 integrated into a MZI} 
\begin{tabular}{c | c | c | c | c | c | c | c}
\hline
 \begin{tabular}{@{}c} $\lambda$ \\ (nm) \end{tabular}& \multicolumn{2}{c|} {\begin{tabular}{@{}c} IL \\(dB) \end{tabular}} &  \multicolumn{2}{c|} {\begin{tabular}{@{}c} ER \\ (dB) \end{tabular}} & \multicolumn{3}{c} {\begin{tabular}{@{}c} $P_\pi$ \\ (mW) \end{tabular}} \\ [0.5ex] 
 \hline
   & Output1 & Output2 & Output1 & Output2 & Output1 & Output2 & Simulation\\
 \hline
 445 & 9.6 & 10.5 & 18.0 & 18.6 & 0.67 & 0.63 & 0.60\\ 
 \hline
 488 & 6.9 & 6.8 & 11.3 & 12.6 & 0.73 & 0.73 & 0.68\\ 
 \hline
 532 & 9.9 & 11.2 & 18.2 & 22.8 & 0.85 & 0.78 & 0.77\\ 
 \hline
 561 & 7.1 & 7.0 & 17.5 & 16.3 & 0.90 & 0.93 & 0.83\\ 
 \hline
\end{tabular}
\end{table}

\section{Appendix}

PS3 had the same edge couplers and MMIs as PS1, but it had 7 SiN waveguides under the TiN heater, which was 10 $\mu$m wide and 983.8 $\mu$m long. The heater resistance used in simulations was 983.8 $\Omega$. The measured value was about 864 $\Omega$. The XTEM of PS3 is shown in Fig. 2(c). The tuning curves of PS3 at 4 different wavelengths are shown in Fig. 6. The IL of PS3 was generally higher than that of PS1 due to the increased waveguide length. The IL, ER, and $P_\pi$ of PS3 are summarized in Table III. The values in Table III are from the first half-period of the MZI tuning. With the same simulation conditions described in Section 2, the simulated $P_\pi$ of PS3 is 0.44 mW at $\lambda = 445$ nm. This is about 32\% lower than the measured value. The discrepancy is likely due to the incomplete removal of Si under the suspended beam. Cross-section images of the suspended beams show that the undercut etch of the Si substrate had an elliptical shape and extended about 8 $\mu$m laterally beyond each edge of the deep trench. The lateral extent of the undercut etch was wide enough to remove all the Si substrate under the suspended beam for PS1, but not directly between the SiO$_2$ anchors for PS3. To model the incomplete undercut etch for PS3, we assumed 2$\mu$m $\times$ 2$\mu$m Si posts connected the suspended beam to the Si substrate at regions directly between the SiO$_2$ anchors. The resulting simulated $P_\pi$ of PS3 is 0.60 mW at 445 nm, close to the measured value. The $P_\pi$ simulations of PS3 with the incomplete undercut are summarized in Table III.



\begin{thebibliography}{10}
\newcommand{\enquote}[1]{``#1''}

\bibitem{Subramanian_2013}
A.~Z. Subramanian, P.~Neutens, A.~Dhakal, R.~Jansen, T.~Claes, X.~Rottenberg,
  F.~Peyskens, S.~Selvaraja, P.~Helin, B.~Du~Bois, K.~Leyssens, S.~Severi,
  P.~Deshpande, R.~Baets, and P.~Van~Dorpe, \enquote{Low-loss singlemode
  \textrm{PECVD} silicon nitride photonic wire waveguides for 532–900 nm
  wavelength window fabricated within a \textrm{CMOS} pilot line,}
  {\protect\textrm{IEEE Photonics Journal}} \textbf{5}, 2202809--2202809
  (2013).

\bibitem{Romero_Garcia_OE_2013}
S.~Romero-Garc\'{i}a, F.~Merget, F.~Zhong, H.~Finkelstein, and J.~Witzens,
  \enquote{Silicon nitride \textrm{CMOS}-compatible platform for integrated
  photonics applications at visible wavelengths,} {\protect\textrm{Opt.
  Express}} \textbf{21}, 14036--14046 (2013).

\bibitem{Domenech_2018}
J.~D. Domenech, M.~A.~G. Porcel, H.~Jans, R.~Hoofman, D.~Geuzebroek, P.~Dumon,
  M.~van~der Vliet, J.~Witzens, E.~Bourguignon, I.~{n}igo Artundo, and
  L.~Lagae, \enquote{Pix4life: photonic integrated circuits for bio-photonics,}
  in \emph{Integrated Photonics Research, Silicon and Nanophotonics,}  (Optical
  Society of America, 2018), p. ITh3B.1.

\bibitem{sacher2019visible}
W.~D. Sacher, X.~Luo, Y.~Yang, F.-D. Chen, T.~Lordello, J.~C.~C. Mak, X.~Liu,
  T.~Hu, T.~Xue, P.~G.-Q. Lo, M.~L. Roukes, and J.~K.~S. Poon,
  \enquote{Visible-light silicon nitride waveguide devices and implantable
  neurophotonic probes on thinned 200 mm silicon wafers,}
  {\protect\textrm{Opt. Express}} \textbf{27}, 37400--37418 (2019).

\bibitem{Al2O3_JSTQE_2019}
C.~Sorace-Agaskar, D.~Kharas, S.~Yegnanarayanan, R.~T. Maxson, G.~N. West,
  W.~Loh, S.~Bramhavar, R.~J. Ram, J.~Chiaverini, J.~Sage, and P.~Juodawlkis,
  \enquote{Versatile silicon nitride and alumina integrated photonic platforms
  for the ultraviolet to short-wave infrared,} {\protect\textrm{IEEE
  Journal of Selected Topics in Quantum Electronics}} \textbf{25}, 1--15
  (2019).

\bibitem{Mohanty_2020}
A.~Mohanty, Q.~Li, M.~A. Tadayon, S.~P. Roberts, G.~R. Bhatt, E.~Shim, X.~Ji,
  J.~Cardenas, S.~A. Miller, A.~Kepecs, and M.~Lipson, \enquote{Reconfigurable
  nanophotonic silicon probes for sub-millisecond deep-brain optical
  stimulation,} {\protect\textrm{Nature Biomedical Engineering}}
  \textbf{4}, 223--231 (2020).

\bibitem{laser_engine_OE_2021}
A.~T. Mashayekh, T.~Klos, D.~Geuzebroek, E.~Klein, T.~Veenstra, M.~Büscher,
  F.~Merget, P.~Leisching, and J.~Witzens, \enquote{Silicon nitride
  \textrm{PIC}-based multi-color laser engines for life science applications,}
  {\protect\textrm{Opt. Express}} \textbf{29}, 8635--8653 (2021).

\bibitem{Sacher_Neurophotonics_2021}
W.~D. Sacher, F.-D. Chen, H.~Moradi-Chameh, X.~Luo, A.~Fomenko, P.~Shah,
  T.~Lordello, X.~Liu, I.~F. Almog, J.~N. Straguzzi, T.~M. Fowler, Y.~Jung,
  T.~Hu, J.~Jeong, A.~M. Lozano, P.~G.-Q. Lo, T.~A. Valiante, L.~C. Moreaux,
  J.~K.~S. Poon, and M.~L. Roukes, \enquote{{Implantable photonic neural probes
  for light-sheet fluorescence brain imaging},}
  {\protect\textrm{Neurophotonics}} \textbf{8}, 1 -- 26 (2021).

\bibitem{poulton2017}
C.~V. Poulton, M.~J. Byrd, M.~Raval, Z.~Su, N.~Li, E.~Timurdogan, D.~Coolbaugh,
  D.~Vermeulen, and M.~R. Watts, \enquote{Large-scale silicon nitride
  nanophotonic phased arrays at infrared and visible wavelengths,}
  {\protect\textrm{Opt. Lett.}} \textbf{42}, 21--24 (2017).

\bibitem{Niffenegger_Nature_2020}
R.~J. Niffenegger, J.~Stuart, C.~Sorace-Agaskar, D.~Kharas, S.~Bramhavar, C.~D.
  Bruzewicz, W.~Loh, R.~T. Maxson, R.~McConnell, D.~Reens, G.~N. West, J.~M.
  Sage, and J.~Chiaverini, \enquote{{Integrated multi-wavelength control of an
  ion qubit},} {\protect\textrm{Nature}} \textbf{586}, 538 -- 542 (2020).

\bibitem{Mehta_Nature_2020}
K.~K. Mehta, C.~Zhang, M.~Malinowski, T.-L. Nguyen, M.~Stadler, and J.~P. Home,
  \enquote{{Integrated optical multi-ion quantum logic},}
  {\protect\textrm{Nature}} \textbf{586}, 533 -- 537 (2020).

\bibitem{west2019low}
G.~N. West, W.~Loh, D.~Kharas, C.~Sorace-Agaskar, K.~K. Mehta, J.~Sage,
  J.~Chiaverini, and R.~J. Ram, \enquote{Low-loss integrated photonics for the
  blue and ultraviolet regime,} {\protect\textrm{APL Photonics}}
  \textbf{4}, 026101 (2019).

\bibitem{timurdogan2019apsuny}
E.~Timurdogan, Z.~Su, R.-J. Shiue, C.~V. Poulton, M.~J. Byrd, S.~Xin, and M.~R.
  Watts, \enquote{\textrm{APSUNY} process design kit (\textrm{PDK}v3. 0): \textrm{O, C and L}
  band silicon photonics component libraries on 300mm wafers,} in \emph{2019
  Optical Fiber Communications Conference and Exhibition (OFC),}  (IEEE, 2019),
  pp. 1--3.

\bibitem{wilmart2019versatile}
Q.~Wilmart, H.~El~Dirani, N.~Tyler, D.~Fowler, S.~Malhouitre, S.~Garcia,
  M.~Casale, S.~Kerdiles, K.~Hassan, C.~Monat \emph{et~al.}, \enquote{A
  versatile silicon-silicon nitride photonics platform for enhanced
  functionalities and applications,} {\protect\textrm{Applied Sciences}}
  \textbf{9}, 255 (2019).

\bibitem{sacher2018monolithically}
W.~D. Sacher, J.~C. Mikkelsen, Y.~Huang, J.~C.~C. Mak, Z.~Yong, X.~Luo, Y.~Li,
  P.~Dumais, J.~Jiang, D.~Goodwill, E.~Bernier, P.~G.-Q. Lo, and J.~K.~S. Poon,
  \enquote{Monolithically integrated multilayer silicon nitride-on-silicon
  waveguide platforms for \textrm{3-D} photonic circuits and devices,}
  {\protect\textrm{Proceedings of the IEEE}} \textbf{106}, 2232--2245
  (2018).

\bibitem{absil2017reliable}
P.~Absil, K.~Croes, A.~Lesniewska, P.~De~Heyn, Y.~Ban, B.~Snyder, J.~De~Coster,
  F.~Fodor, V.~Simons, S.~Balakrishnan \emph{et~al.}, \enquote{Reliable 50\textrm{Gb/s}
  silicon photonics platform for next-generation data center optical
  interconnects,} in \emph{2017 IEEE International Electron Devices Meeting
  (IEDM),}  (IEEE, 2017), pp. 34--2.

\bibitem{towerjazz}
{Tower Semiconductor},
  \url{https://towersemi.com/manufacturing/mpw-shuttle-program/}.

\bibitem{arbabi2013measurements}
A.~Arbabi and L.~L. Goddard, \enquote{Measurements of the refractive indices
  and thermo-optic coefficients of $\mathrm{Si_3 N_4}$ and $\mathrm{SiO_x}$
  using microring resonances,} {\protect\textrm{Optics Letters}}
  \textbf{38}, 3878--3881 (2013).

\bibitem{Liang_CLEO_2019}
G.~Liang, H.~Huang, S.~Shrestha, A.~Mohanty, X.~Ji, M.~C. Shin, M.~Lipson, and
  N.~Yu, \enquote{Micron-scale, efficient, robust phase modulators in the
  visible,} in \emph{Conference on Lasers and Electro-Optics, OSA Technical
  Digest (Optical Society of America, 2019),}  (2019), p. JTh5B.4.

\bibitem{Huang_CLEO_2020}
H.~Huang, G.~Liang, A.~Mohanty, X.~Ji, M.~C. Shin, M.~Lipson, and N.~Yu,
  \enquote{Robust miniature pure-phase modulators at k = 488 nm,} in
  \emph{Conference on Lasers and Electro-Optics, OSA Technical Digest (Optical
  Society of America, 2020),}  (2020), p. STh1J.4.
  
\bibitem{yu2021micron}
  N.~Yu and G.~Liang and H.~Huang and A.~Mohanty, X.~Ji, M.~Shin, M.~Carter, S.~Shrestha, M.~Lipson,
  \enquote{Micron-scale, efficient, and robust phase modulators at visible wavelengths,}
  \emph{Proc. SPIE 11694, Photonic and Phononic Properties of Engineered Nanostructures XI,} 1169407, (2021).


\bibitem{notaros2018integrated}
M.~Notaros, M.~Raval, J.~Notaros, and M.~R. Watts, \enquote{Integrated
  visible-light liquid-crystal phase modulator,} in \emph{Frontiers in Optics,}
   (Optical Society of America, 2018), pp. FW6B--5.

\bibitem{notaros2019liquid}
J.~Notaros, M.~Notaros, M.~Raval, and M.~R. Watts,
  \enquote{Liquid-crystal-based visible-light integrated optical phased
  arrays,} in \emph{2019 Conference on Lasers and Electro-Optics (CLEO),}
  (IEEE, 2019), pp. 1--2.

\bibitem{desiatov2019ultra}
B.~Desiatov, A.~Shams-Ansari, M.~Zhang, C.~Wang, and M.~Lon{\v{c}}ar,
  \enquote{Ultra-low-loss integrated visible photonics using thin-film lithium
  niobate,} {\protect\textrm{Optica}} \textbf{6}, 380--384 (2019).

\bibitem{Kasahara_ECOC_2008}
R.~Kasahara, K.~Watanabe, M.~Itoh, Y.~Inoue, and A.~Kaneko, \enquote{Extremely
  low power consumption thermooptic switch (0.6 \textrm{mW}) with suspended
  ridge and silicon-silica hybrid waveguide structures,} in \emph{2008 34th
  European Conference on Optical Communication,}  (2008), pp. 1--2.

\bibitem{Densmore_OE_2009}
A.~Densmore, S.~Janz, R.~Ma, J.~H. Schmid, D.-X. Xu, A.~Delâge, J.~Lapointe,
  M.~Vachon, and P.~Cheben, \enquote{Compact and low power thermo-optic switch
  using folded silicon waveguides,} {\protect\textrm{Opt. Express}}
  \textbf{17}, 10457--10465 (2009).

\bibitem{fang2011ultralow}
Q.~Fang, J.~F. Song, T.-Y. Liow, H.~Cai, M.~B. Yu, G.~Q. Lo, and D.-L. Kwong,
  \enquote{Ultralow power silicon photonics thermo-optic switch with suspended
  phase arms,} {\protect\textrm{IEEE Photonics Technology Letters}}
  \textbf{23}, 525--527 (2011).

\bibitem{lu2015michelson}
Z.~Lu, K.~Murray, H.~Jayatilleka, and L.~Chrostowski, \enquote{Michelson
  interferometer thermo-optic switch on \textrm{SOI} with a 50-$\mathrm{\mu W}$ power
  consumption,} {\protect\textrm{IEEE Photonics Technology Letters}}
  \textbf{27}, 2319--2322 (2015).

\bibitem{MurrayOE2015}
K.~Murray, Z.~Lu, H.~Jayatilleka, and L.~Chrostowski, \enquote{Dense dissimilar
  waveguide routing for highly efficient thermo-optic switches on silicon,}
  {\protect\textrm{Opt. Express}} \textbf{23}, 19575--19585 (2015).

\bibitem{celo2016thermo}
D.~Celo, D.~J. Goodwill, J.~Jiang, P.~Dumais, M.~Li, and E.~Bernier,
  \enquote{Thermo-optic silicon photonics with low power and extreme resilience
  to over-drive,} in \emph{2016 IEEE Optical Interconnects Conference (OI),}
  (IEEE, 2016), pp. 26--27.

\bibitem{yiding2021ec}
Y.~Lin, J.~C.~C. Mak, H.~Chen, X.~Mu, A.~Stalmashonak, Y.~Jung, X.~Luo,
  P.~G.-Q. Lo, W.~D. Sacher, and J.~K.~S. Poon, \enquote{Low-loss broadband
  bi-layer edge couplers for visible light,} {\protect\textrm{Opt.
  Express}} \textbf{29}, 34565--34576 (2021).

\bibitem{yiding2021pd}
Y.~Lin, Z.~Yong, X.~Luo, P.~G.-Q. Lo, W.~D. Sacher, and J.~K. Poon,
  \enquote{Silicon nitride waveguide-integrated silicon photodiodes for blue
  light,} in \emph{CLEO: Science and Innovations,}  (Optical Society of
  America, 2021), p. SM1A.3.

\bibitem{Elshaari_IEEE_Photonics_Journal_2016}
A.~W. Elshaari, I.~E. Zadeh, K.~D. Jöns, and V.~Zwiller, \enquote{Thermo-optic
  characterization of silicon nitride resonators for cryogenic photonic
  circuits,} {\protect\textrm{IEEE Photonics Journal}} \textbf{8}, 1--9
  (2016).

\end{thebibliography}
\end{document}